\documentclass{article}
\usepackage{arxiv}
\usepackage{graphicx} 

\usepackage{amsmath}
\usepackage{amssymb}
\usepackage{bm}
\usepackage{mathtools}
\usepackage{booktabs}
\usepackage{xcolor}
\usepackage{subcaption}
\usepackage{comment}
\usepackage{physics}
\usepackage{subcaption}
\usepackage{multirow}

\usepackage[algo2e,ruled,vlined]{algorithm2e}
\DontPrintSemicolon

\newcommand{\bs}[1]{\boldsymbol{#1}}
\newcommand{\bsigma}{\bs{\sigma}}
\newcommand{\cauchy}{\bsigma}
\newcommand{\devcauchy}{\bs{s}}
\newcommand{\boldI}{\bs{I}}
\newcommand{\pdir}{\tilde{\bs{e}}}
\newcommand{\pval}{\tilde{\sigma}}
\newcommand{\devpval}{\tilde{s}}

\newcommand{\cref}[1]{Ref.~\cite{#1}}

\title{Thermodynamically Consistent Hybrid and Permutation-Invariant Neural Yield Functions for Anisotropic Plasticity}

\renewcommand{\shorttitle}{Neural Yield Functions for Anisotropic Plasticity}

\author{
Asghar A. Jadoon, \\
{\it
Department of Aerospace Engineering \& Engineering Mechanics}, \\
{\it
The University of Texas at Austin}, \\
Austin, TX 78712
\And
Ravi G. Patel, \\
{\it Sandia National Laboratories},\\
Albuquerque, NM 87185
\And
Brian N. Granzow, \\
{\it Sandia National Laboratories},\\
Albuquerque, NM 87185
\And
Reese E. Jones, \\
{\it Sandia National Laboratories},\\
Livermore, CA 94551
\And
D. Thomas Seidl, \\
{\it Sandia National Laboratories},\\
Albuquerque, NM 87185
\And
Jan N. Fuhg\thanks{{correspondence: \tt jan.fuhg@utexas.edu}}, \\
{\it
Department of Aerospace Engineering \& Engineering Mechanics}, \\
{\it
\& The Oden Institute of Computational Science and Engineering,} \\
{\it
The University of Texas at Austin}, \\
Austin, TX 78712
}

\date{\today}

\begin{document}

\maketitle

\section*{Abstract}
Plastic anisotropy in metals remains challenging to model. This is partly because conventional phenomenological yield criteria struggle to combine a highly descriptive, flexible representation with constraints, such as convexity, dictated by thermodynamic consistency. To address this gap, we employ architecturally-constrained neural networks and develop two data-driven frameworks: (i) a hybrid model that augments the Hill yield criterion with an Input Convex Neural Network (ICNN) to get an anisotropic yield function representation in the six-dimensional stress space and (ii) a permutation-invariant input convex neural network (PI-ICNN) that learns an isotropic yield function representation in the principal stress space and embeds anisotropy through one ($\text{PI-ICNN}_1$) or two ($\text{PI-ICNN}_2$) linear stress transformations. We calibrate the proposed frameworks on a sparse Al-7079 extrusion experimental dataset comprising 12 uniaxial samples with measured yield stresses and Lankford ratios. To test the robustness of each framework, nine datasets were generated using k-fold cross-validation. These datasets were then used to quantitatively compare Hill-48, Yld2004-18p, pure ICNNs, the hybrid approach, and the two PI-ICNN frameworks. While ICNNs and hybrid approaches can almost perfectly fit the training data, they exhibit significant over-fitting, resulting in high validation and test losses. In contrast, both PI-ICNN frameworks demonstrate better generalization capabilities, with $\text{PI-ICNN}_1$ even outperforming Yld2004-18p on the validation and test data, despite using only a single linear stress transformation, whereas Yld2004-18p employs two. These results demonstrate that PI-ICNNs unify physics-based constraints with the flexibility of neural networks, enabling the accurate prediction of both yield loci and Lankford ratios from minimal data. The approach opens a path toward rapid, thermodynamically consistent constitutive models for advanced forming simulations and future exploration of coupled hardening or microstructure‑informed design.

\section{Introduction}

Although mechanical anisotropy of \textit{elastic} materials was already being explored in the middle of the 19\textsuperscript{th} century \cite{helbig200575}. Tresca \cite{tresca1864} was the first to formally attempt to define a material-specific yield condition in 1864 \cite{koiter1960general}. 
Later, von Mises \cite{vonmises1913} and Drucker \cite{drucker1959} extended this work by developing new yield criteria and by refining the theory of rate-independent plasticity.  However, in these initial works, the yield function was assumed to be isotropic. In 1948, Hill \cite{hill1948theory}   introduced the first generally applicable theory of anisotropic yield, laying the foundation for modern formulations of plasticity in directionally dependent materials. 
Hill derived an anisotropic yield criterion that was quadratic in nature (similar to von Mises yield criterion) but with an additional 6 parameters to specify the state of anisotropy. Since then, a large number of phenomenological laws for anisotropic yield functions have been proposed.

For example, Hosford \cite{Hosford1972} presented his generalized yield criterion in 1972.  He introduced a parameter in the yield criterion to model a range of isotropic yield criteria. Hill \cite{hill1979theoretical}  extended this idea to anisotropic yield and defined a generalized Hill criterion which was no longer constrained to be quadratic. Barlat et al. \cite{barlat1987prediction}  further generalized Hill's criterion by introducing new stress tensor invariants for planar anisotropy. This was later extended to a three-dimensional space \cite{BARLAT1991693} which, similar to \cref{karafillis1993general}, used a linear transformation of the Cauchy stress tensor to introduce anisotropy. This concept introduced a simple framework to model anisotropy because it relied on existing isotropic yield function representations. The affine transformation made it straightforward to design anisotropic yield functions that are convex. Later on, to increase their expressivity, the use of two linear transformations was explored \cite{barlat2003plane, bron2004yield, BARLAT20051009, BARLAT2007876} which resulted in a better representation of the yield surface at the cost of additional parameters. This approach has also recently been extended in other notable works \cite{cazacu2006orthotropic, yoshida2013user, yoon2006prediction, yoon2014asymmetric, plunkett2008orthotropic, aretz2013new, lou2018anisotropic}.

We hypothesize that the need for additional transformations is partly due to the fact that the isotropic functions that are used in these models are not sufficiently representative. 
In particular, the isotropic yield functions in these references are commonly formulated as power-law functions of stress differences or transformed stress components. This structure is greatly motivated by the need to preserve the convexity of the yield surface, which is essential for both physical admissibility and numerical stability \cite{jirasek2001inelastic, simo2006computational}. However, power-law formulations are limited in their expressive capacity.
To address this, recent advances have turned to machine learning regression techniques that offer greater representative power \cite{fuhg2024review}.
The application of machine learning to plasticity modeling dates back to the early 1990s \cite{ghaboussi1991knowledge, ghaboussi1998new}, and the field has since seen significant advancements. For a comprehensive review of data-driven approaches in modeling plasticity, the reader is referred to Section 5.1 of \cref{fuhg2024review}. In this work, we focus specifically on contributions related to data-driven anisotropic yield modeling and those that directly motivate the methodologies proposed herein. Building on the aforementioned classical works in data-driven plasticity modeling, different neural network architectures have been employed to model plastic behavior \cite{mozaffar2019deep, huang2020machine, maurizi2022predicting, ibragimova2022convolutional} without necessarily relying on the well-established continuum plasticity concepts of yield surfaces, flow rules, etc. Instead, the aforementioned works aim to directly establish a stress-strain map and thus do not make a distinction between isotropic or anisotropic yield behavior. In contrast, other works have relied on machine learning techniques to model anisotropic yield surfaces that can easily be integrated into frameworks based on continuum elastoplasticity. For instance, Support Vector Classification was explored in \cref{hartmaier2020data} to act as a surrogate for anisotropic yield function. However, the convexity of this surrogate was not explicitly considered. In \cref{flaschel2022discovering}, symbolic regression is used to model yield surfaces with built-in convexity through sparsification of a chosen ansatz. Neural networks have also been employed for modeling yield surfaces with earlier works \cite{vlassis2021sobolev} enforcing convexity through a penalization term in the loss function. Recent approaches \cite{fuhg2022machine,soare2025use} have, however, moved towards the use of Input Convex Neural Networks (ICNNs) \cite{amos2017input} with convexity constraints built into the neural network architecture.

While neural networks have been employed to good use for modeling the yield behavior, virtually all published architectures take inputs from the six-dimensional stress space (or 3 dimensions in 2D). However, to reduce the input space, classical phenomenological models have been developed that use the three-dimensional principal stress space as input, see e.g. Refs \cite{barlat1987prediction,karafillis1993general}.
This is difficult to model for neural networks because conventional feed-forward neural networks (FFNNs) or their convexity-constrained variants, e.g, ICNNs, are not intrinsically permutation-invariant, a property essential for preserving convexity (and coordinate invariance) and consequently, thermodynamic consistency in principal stress space. In this work, we address this gap by designing a neural network architecture that is both permutation-invariant and convex with respect to the principal stresses. These networks were recently used in the context of data-driven constitutive modeling in \cref{TEPOLE2025113469}, where the authors built on the permutation-invariant neural networks proposed in \cref{kimura2024permutation} based on Deep Sets \cite{zaheer2017deep} and constrained them to have built-in convexity in the architecture. The authors used these networks to model hyperelastic strain energies in terms of principal stretches. In this work, we use a similar architecture to design Permutation-Invariant Input Convex Neural Networks (PI-ICNNs) to model isotropic yield surfaces in terms of principal stresses, and then use linear transformations to incorporate anisotropy into the models.

Another approach, namely a hybrid or model-data-driven approach, relies on the correction of yield surfaces predicted by a phenomenological model through data-driven techniques. To the best of the authors' knowledge, the use of data-driven correction to phenomenological anisotropic yield functions was first introduced in \cref{ibanez2019hybrid}, where the authors used a nonlinear sparse identification technique to formulate the correction term for Yld2004-18p \cite{BARLAT20051009}. This approach was later extended in \cref{FUHG2023104925} where the authors used other techniques such as support vector regression, Gaussian regression, and input convex neural networks to construct the error or correction term for the anisotropic yield model of \cref{cazacu2004criterion} in 2D and with synthetic data. It was shown that, irrespective of the regression technique used for the correction, these models could fit the desired yield surfaces. In contrast to \cref{ibanez2019hybrid, FUHG2023104925}, which applied corrections to an already anisotropic yield function, i.e., Hill's yield criterion, \cref{ghnatios2024new} chose to start with an isotropic von Mises yield function to successfully reconstruct an anisotropic yield function proposed in \cref{cazacu2006orthotropic}. However, all the aforementioned works focus only on fitting the yield function and do not take into consideration the Lankford ratios. Furthermore, they have not been used on experimental data.
In this work, we perform the convex correction to not only find a better for for the yield values but also for the Lankford coefficients and test it on experimental data.

To summarize, in this work, we present two distinct methodologies for modeling anisotropic yield functions. The first approach can be classified as a model-data-driven approach, sharing conceptual foundations with Refs. \cite{ibanez2019hybrid, FUHG2023104925}. In this approach, we formulate a hybrid yield function in the six-dimensional stress space by first training Hill's yield criterion \cite{hill1948theory} on the dataset and then using an ICNN to enforce convex corrections to the discrepancy in Hill's predictions. The second approach is a purely data-driven one where we formulate the yield function in principal stress space using PI-ICNNs and use linear transformations of stresses to model anisotropic yield. Both these approaches are tested on the dataset presented in \cref{corona2021anisotropic}, and a comparative analysis is performed between the two approaches as well as with Hill's criterion, Yld2004-18p \cite{BARLAT20051009}, and pure ICNNs. This paper is organized as follows: A theoretical framework is laid out in Section 2, which summarizes the necessary concepts before all the approaches employed in this study are detailed in Section 3. The dataset used for modeling the yield functions, along with the calibration procedures, is explained in Section 4. The results are presented in Section 5 before concluding the work in Section 6.

\section{Theory}
We investigate an elastoplastic material response and assume a classical additive split of the infinitesimal strain into elastic and plastic components  
\begin{equation}
\bm{\epsilon} = \bm{\epsilon}^{e} + \bm{\epsilon}^{p} \ .
\end{equation}
Furthermore, we assume associative plasticity with isotropic hardening and postulate the existence of a free energy function $\Psi(\bm{\epsilon}^{e}, \kappa)$ that is dependent on the scalar hardening variable $\kappa$.
The plastic behavior is then modeled with a yield function $f(\bm{\sigma},k)$ that is dependent on the stress $\bm{\sigma} = \partial_{\bm{\epsilon}^{e}} \Psi$ and the thermodynamic force $k = \partial_{\kappa} \Psi$. 
For linear elasticity, the free energy $\Psi$ is quadratic in the elastic strain with a constant fourth-order tangent modulus or stiffness tensor $\mathbb{C}$. This defines the stress
\begin{equation}
    \bm{\sigma} = \mathbb{C}:\bm{\epsilon}^e = \mathbb{C}:[\bm{\epsilon} - \bm{\epsilon}^p]
\end{equation}
The yield function is then classically split into
\begin{equation}
    f(\bm{\sigma}, k) = \Phi(\bm{\sigma}) - Y_0 - \varphi(k) \ ,
    \label{eq:Yield_general}
\end{equation}
where $\Phi$ is the equivalent or effective stress function, $Y_0$ is the initial yield stress and $\varphi(k)$ is a hardening function. The material deforms elastically when $f<0$, and $f=0$ defines the onset of plasticity/plastic loading.

The evolution of plastic strains is then defined by an associative flow rule, i.e.,
\begin{equation}
    \dot{\bm{\epsilon}}^p = \dot{\lambda} \frac{\partial f}{\partial\bm{\sigma}} \ ,
\end{equation}
where $\lambda$ is a plastic multiplier given by the Karush-Kuhn-Tucker conditions
\begin{equation} \label{KKT}
f \leq 0 \ , \quad \dot{\lambda}\geq 0, \quad f \dot{\lambda}= 0 \, .
\end{equation}
For sheet metals, it is often useful (see e.g. \cref{suzuki1987relationship, rickhey2021evolution}) to quantify the resistance of the sheet to through-thickness thinning during in-plane stretching. The plastic strain ratio, commonly called the r-value or Lankford coefficient, is defined in uniaxial tension along the y-axis as
\begin{equation}
r = \frac{\bm{\epsilon}^p_z}{\bm{\epsilon}^p_x}    \ .
\end{equation}
Here, $\bm{\epsilon}^p_z$ is the plastic strain transverse to the loading direction and $\bm{\epsilon}^p_x$ is the through-thickness plastic strain. This coefficient is a quantitative measure of the plastic anisotropy of rolled sheet metal. In particular, plastic isotropy is characterized by $r=1$, whereas we observe different r-values for different orientations in the case of plastic anisotropy. For an accurate prediction of the elastoplastic behavior of a material exhibiting plastic anisotropy, both the yield surface and the r-values must be matched in calibration.

For thermodynamic consistency, the yield function is a convex function of the stress tensor $\bm{\sigma}$, see Appendix \ref{sec:Thermodynam}. For a yield function of the form in Eq. \eqref{eq:Yield_general}, this requirement is interchangeable with the convexity of the equivalent stress function $\Phi$. Due to the symmetry of the stress tensor, this requires convexity in the six-dimensional stress space, which we define as
\begin{equation}
    \bar{\bm{\sigma}} = \{\sigma_{11}, \sigma_{22}, \sigma_{33}, \sigma_{12}, \sigma_{23}, \sigma_{13} \}^T .
\end{equation}
However, for classical isotropic yielding, we can formulate the yield or the equivalent stress function in the three-dimensional principal stress space ($\sigma_{1},\sigma_{2},\sigma_{3}$) using the isotropic function:
\begin{equation}
    f(\bm{\sigma}, k) = g(\sigma_{1},\sigma_{2},\sigma_{3}, k) \, .
\end{equation}
 To ensure convexity in the principal stress space, $ g$ is required to be convex and permutation-invariant in the principal stresses
\cite{davis1957all}. Furthermore, if the plastic deformation is pressure-independent, the point of yield is only dependent on the deviatoric stress $\bm{s} = \bm{\sigma} - (1/3)\text{trace}(\bm{\sigma})\bm{I}$, which then gives:
\begin{equation}
    f(\bm{\sigma}, k) = F(\bm{s}, k) = G(s_{1},s_{2},s_{3}, k) = g(\sigma_{1},\sigma_{2},\sigma_{3}, k),
\end{equation}
where $s_1$, $s_2$ and $s_3$ are the principal deviatoric stresses. To reintroduce anisotropy into the isotropic yield function, we follow Barlat et al. \cite{BARLAT20051009}, and postulate that the stress is altered by a linear transformation $\bm{L}$ such that
\begin{equation}
    {\bm{s}'} = \bm{C}\bar{\bm{s}} = \bm{C}\bm{T}\bar{\bm{\sigma}} = \bm{L}\bar{\bm{\sigma}} \ ,
\end{equation}
Here, $\bar{\bm{s}} =\{s_{11}, s_{22}, s_{33}, s_{12}, s_{23}, s_{13} \}^T$, $\bm{C}$ contains a set of anisotropy coefficients and $\bm{T}$ transforms the stress to its deviator:
\begin{equation}
\bm{C} \;=\;
\begin{bmatrix}
 0 & -c_{12} & -c_{13} & 0 & 0 & 0 \\
 -c_{21} & 0 & -c_{23} & 0 & 0 & 0 \\
 -c_{31} & -c_{32} & 0 & 0 & 0 & 0 \\
 0 & 0 & 0 & c_{44} & 0 & 0 \\
 0 & 0 & 0 & 0 & c_{55} & 0 \\
 0 & 0 & 0 & 0 & 0 & c_{66}
\end{bmatrix},
\qquad
\bm{T} \;=\; \frac{1}{3}
\begin{bmatrix}
 2 & -1 & -1 & 0 & 0 & 0 \\
 -1 & 2 & -1 & 0 & 0 & 0 \\
 -1 & -1 & 2 & 0 & 0 & 0 \\
 0 & 0 & 0 & 3 & 0 & 0 \\
 0 & 0 & 0 & 0 & 3 & 0 \\
 0 & 0 & 0 & 0 & 0 & 3
\end{bmatrix}
.
\label{eq:CT_matrices}
\end{equation}
using the Voigt map between the components of a 4th order tensor in 3-space and a 2nd 6$\times$6 matrix. 
Note $\bm{C}$ is effectively a parameterized structure tensor characterizing symmetries from isotropic to orthorhombic and $\bm{T}$ is the projector from symmetric onto deviatoric 2nd order tensors; hence, $\bm{C}$ is used to represent material symmetries and $\bm{T}$ is employed to preserve plastic incompressibility.






We combine them and define the linear transformation operator $\bm{L}=\bm{C}\bm{T}$ as:
\begin{equation}
\bm{L} \;=\;
\frac{1}{3}
\begin{bmatrix}
 c_{12}+c_{13} & -2c_{12}+c_{13} &  c_{12}-2c_{13} & 0      & 0      & 0 \\
-2c_{21}+c_{23} &  c_{21}+c_{23} &  c_{21}-2c_{23} & 0      & 0      & 0 \\
-2c_{31}+c_{32} &  c_{31}-2c_{32} &  c_{31}+c_{32} & 0      & 0      & 0 \\
0               & 0               & 0               & 3c_{44} & 0      & 0 \\
0               & 0               & 0               & 0      & 3c_{55} & 0 \\
0               & 0               & 0               & 0      & 0      & 3c_{66}
\end{bmatrix}
\label{eq:projection_L}
\end{equation}
We can then write
\begin{equation}
    f(\bm{\sigma}, k) = h({s}'_{1},{s}'_{2},{s}'_{3}, k) \ ,
\end{equation}
where ${s}'_i$ with $i=1,2,3$ are the eigenvalues of the tensorial form of $\bm{{s}}'$ and we require $h$ to be a convex and permutation-invariant function in the principal deviatoric stresses to be thermodynamically consistent.

\section{Methods} \label{Sec:Methods}
In this section, we describe the approaches employed in this study to model anisotropic yield surfaces. Starting with some well-known phenomenological models, we move on to NN-based approaches, where we first outline a framework for constructing Input Convex Neural Networks (ICNNs). These ICNNs are then employed alongside phenomenological yield functions to formulate a model-data-driven or hybrid framework. Finally, we introduce the architecture for Permutation-Invariant Input Convex Neural Networks (PI-ICNNs). 

In the following, we assume tension-compression symmetry. This requires the yield function to be an even function, that is, $f(\bm{\sigma})=f(-\bm{\sigma})$. While phenomenological functions can be chosen to directly satisfy this requirement, neural networks do not intrinsically have this property. However, any function $\psi(x)$ can be made even through the relation
\begin{equation}
    \tilde\psi(x) = \frac{\psi(x) + \psi(-x)}{2} \, ,
\end{equation}
where $\tilde{\psi}(x)$ denote the even part of $\psi(x)$. The same transformation can be applied to obtain even symmetric neural networks. In the case of tension-compression asymmetry, this transformation can be omitted.

An additional requirement for yield functions of the form of Eq. \eqref{eq:Yield_general} is that of positive homogeneity of degree one (so that the effective stress is norm-like). It derives from the assumption of isotropic hardening, i.e., the yield function should preserve its initial shape when it expands under isotropic hardening. Again, phenomenological functions have been proposed that fulfill this requirement, albeit at the expense of representative capabilities. Neural networks, in general, are not positive homogeneous and consequently do not satisfy this constraint. Instead of letting the neural network learn this behavior, we handle this by introducing a variable $\beta$ such that, ignoring isotropic hardening, $f(\beta\bm{\sigma})$ is always zero, i.e.,
\begin{equation}
    f(\beta \bm{\sigma}) = \Phi(\beta \bm{\sigma}) - Y_0 \overset{!}{=} 0 \ ,
    \label{eq:scaling_general}
\end{equation}
and for any given stress $\bm{\sigma}$, we find $\beta$ that satisfies Eq. \eqref{eq:scaling_general} using a root-finding algorithm. This treatment is particularly effective in the low-data regime where hardening behavior is undercharacterized or unprobed. Clearly, if the stresses originally lie inside the yield surface and the material exhibits elastic response, $\beta$ will always be greater than 1 so as to expand the stresses to get to the yield surface. Conversely, if the stresses lie outside the original yield surface, $\beta$ must be less than one to bring them back to the yield surface. Using this, we modify the original form of the yield function such that we now have:
\begin{equation}
\tilde{f}(\beta \bm{\sigma}) = \frac{\Phi(\beta \bm{\sigma})}{\beta} - Y_0 \ .
\end{equation}
Since the numerator in the first term will always be equal to $Y_0$ due to the $\beta$ scaling, the term itself will be:
\begin{equation}
\frac{\Phi(\beta \bm{\sigma})}{\beta} 
\begin{cases}
< Y_0, & \text{if } \beta > 1 \quad \text{(elastic)} \\
> Y_0, & \text{if } \beta < 1 \quad \text{(plastic)}
\end{cases}
\end{equation}
This results in a shape-consistent scaling of the yield function as desired for isotropic hardening. Since tension-compression symmetry and positive homogeneity are dealt with in a rather general manner, they are detailed here. However, thermodynamic consistency or convexity of the yield surface is more framework-specific and will be discussed along with the approaches used in this work in the following sections.

\subsection{Phenomenological models}
Two well-established phenomenological, anisotropic yield criteria are adopted in the present study: 
\begin{enumerate}
    \item The quadratic Hill model \cite{hill1948theory}, hereafter Hill-48,
\begin{equation}
    \Phi(\bm\sigma) = \left[ F(\sigma_{22}-\sigma_{33})^2 + G(\sigma_{33}-\sigma_{11})^2 + H(\sigma_{11}-\sigma_{22})^2 + 2L\sigma_{23}^2 + 2M\sigma_{31}^2 + 2N\sigma_{12}^2 \right]^{\frac{1}{2}}  , 
    \label{Hill-48}
\end{equation}
where $F, G, H, L, M, N$ are the anisotropy coefficients, and the generalization of the quadratic form of a J2 model is evident.
\item 
 The 18-parameter formulation of Barlat et al. \cite{BARLAT20051009}, hereafter Yld2004-18p,
\begin{equation}
\begin{aligned}
\Phi(\bm\sigma) = &\Biggl[ \frac{1}{4}(\lvert {s}'_{1}-{s}''_{1}\rvert^{a}
 +\lvert {s}'_{1}-{s}''_{2}\rvert^{a}
 +\lvert {s}'_{1}-{s}''_{3}\rvert^{a}
 +\lvert {s}'_{2}-{s}''_{1}\rvert^{a}
 +\lvert {s}'_{2}-{s}''_{2}\rvert^{a} \\
 &+\lvert {s}'_{2}-{s}''_{3}\rvert^{a}
 +\lvert {s}'_{3}-{s}''_{1}\rvert^{a}
 +\lvert {s}'_{3}-{s}''_{2}\rvert^{a}
 +\lvert {s}'_{3}-{s}''_{3}\rvert^{a}) \Biggr]^ \frac{1}{a}\, .
\end{aligned}
\label{eq:Yld2004-18p}
\end{equation}  
The transformed stresses $s'_i$ and $s''_i$ are obtained through two linear transformations of the stress tensor as:
\begin{equation}\label{eq:Lineartransformations}
    \bm{s}' = \bm{L}'\bar{\bm{\sigma}}, \qquad \bm{s}'' = \bm{L}''\bar{\bm{\sigma}} ,
\end{equation}
where $\bm{L}'$ and $\bm{L}''$ have the forms as Eq. \eqref{eq:projection_L} but independent coefficients.  
\end{enumerate}
As discussed in the Introduction, the need for more than one transformation stems from the restricted representative power of Eq. \eqref{eq:Yld2004-18p}. We remark that both Hill-48 and Yld2004-18p are tension-compression symmetric. 
Apart from the way anisotropy is introduced, they are also formulated in different stress spaces. In particular, Hill-48 is directly expressed in the six-dimensional Cauchy stress and is therefore thermodynamically consistent by being convex in terms of these stress components. Due to the quadratic nature of Hill-48, positivity of the anisotropic coefficients is a sufficient condition for convexity of the yield function. For a more detailed insight into the necessary and sufficient conditions for convexity, as well as other constraints restricting the degree of anisotropy, interested readers are referred to \cref{ottosen2005mechanics}. In contrast, Yld2004-18p is defined in terms of deviatoric principal stresses obtained from two linear stress transformations. Therefore, both convexity and permutation invariance in these arguments are required. Yld2004-18p is convex for $a>1$ and permutation-invariant since it is form-invariant.


\subsection{Input convex neural networks}
We first consider modeling the yield function with a purely data-driven approach employing neural networks. In particular, if we set aside the hardening variable, we are approximating the pressure-independent yield function as the output of a neural network
\begin{equation}
    f(\bm{\sigma}) = f_{\text{NN}} \left(\bar{\bm{s}} \right)
\end{equation}
that takes the deviatoric Cauchy stresses $\bar{\bm{s}}$ in vectorized form as its input.
Thermodynamic consistency requires the output of the neural network to be convex in its input arguments. A naive approach to enforce this would be to add an additional term in the loss function of the neural network, which penalizes the network for violating convexity. However, this runs the risk of the network being non-convex for unseen data. Therefore, we instead use Input Convex Neural Networks (ICNNs) \cite{amos2017input} that have built-in architectural constraints that ensure the network is always convex in its inputs. ICNNs are feed-forward neural networks with pass-through layers, positivity constraints on some of the weights, and monotone activations. 
The scalar-valued output of an ICNN with input $\mathbf{x}_0$ and $k$ layers can be obtained using the update rule
\begin{equation}
\begin{aligned}
        \mathbf{x}_1 &= \Theta \left( \mathbf{W}_0^{[xx]} \mathbf{x}_0 + \mathbf{b}_0 \right) \\
        \mathbf{x}_{h+1} &= \Theta \left( \mathbf{W}_h^{[xx]} \mathbf{x}_h + \mathbf{W}_h^{[xx_0]} \mathbf{x}_0 + \mathbf{b}_h \right) , \quad h = 1, \ldots, k-1 \ . \label{eq:ICNN}
\end{aligned}
\end{equation}
Here, the weights $\mathbf{W}_{0:k-1}^{[xx]}$ and $\mathbf{W}_{1:k-1}^{[xx_0]}$, along with biases $\mathbf{b}_{0:k-1}$ are the trainable parameters and $\Theta$ denotes a nonlinear activation function. For the network to be convex, the weights $\mathbf{W}_{1:k-1}^{[xx]}$ are constrained to be non-negative, and $\Theta$ must be a convex, monotonically non-decreasing function \cite{amos2017input}. Throughout this work, unless otherwise stated, we use the Softplus activation function \cite{fuhg2022machine}. 
For pressure-independent yielding, this allows us to model the yield function as an ICNN
\begin{equation}
    f(\bm{\sigma}) = f_{\text{ICNN}} \left(\bar{\bm{s}} \right)
\end{equation}
which in this form guarantees thermodynamic consistency.

\subsection{Hybrid Modeling Approach}
The hybrid modeling approach explored in this work is grounded in the principle of enhancing classical phenomenological yield functions using data-driven correction mechanisms while preserving thermodynamic consistency \cite{FUHG2023104925}. In general, this framework allows for the convex correction of \emph{any} existing yield function, phenomenological or otherwise, by augmenting it with a learnable correction term implemented through an ICNN. The role of the ICNN is to account for systematic discrepancies (model-form error) between the model predictions and experimental observations, while ensuring that the resulting corrected yield function remains convex and thus thermodynamically admissible.

 Let $\Phi_{\text{base}}$ represent the equivalent stress function of a given base model and $\Phi_{\text{ICNN}}$ denote the data-driven correction term learned by the ICNN. The hybrid equivalent stress function is then defined as:
\begin{equation}
\Phi_{\text{Hybrid}}(\bm{\sigma}) = \Phi_{\text{base}}(\bar{\bm{s}}) + \Phi_{\text{ICNN}}\left(\bar{\bm{s}} \right).
\end{equation}
In the present work, we take $\Phi_{\text{base}} \equiv \Phi_{\text{Hill\text{-}48}}$ as a demonstration case. Since both components $\Phi_{\text{base}}$ and $\Phi_{\text{ICNN}}$ are convex by construction, their sum is also convex~\cite{boyd2004convex}, and the resulting yield function remains thermodynamically consistent.

The training procedure is carried out in two sequential stages. First, the parameters of the base model (e.g., Hill-48 anisotropy coefficients) are calibrated using available experimental data to obtain an initial best-fit estimate of the yield surface. Then, an ICNN is trained to learn the residual error between the base model prediction and the ground truth.

This approach offers two key advantages:
\begin{enumerate}
    \item It enables the reuse and refinement of well-understood models with accepted parameterizations, incorporating physical intuition while expanding expressivity.
    \item It guarantees thermodynamic consistency by enforcing convexity at the architectural level of the neural network.
\end{enumerate}

We remark that while Hill-48 is used in this work due to its simplicity and widespread use in modeling anisotropic sheet metals, the methodology is general and can be extended to more complex or nonlinear yield functions in other studies.

\subsection{Permutation-invariant input convex neural networks}
Formulating anisotropic yield functions in terms of principal stresses and linear transformations, as done in \cref{BARLAT1991693}, halves the number of independent stress components. 
However, so far, the presented convex neural network architectures cannot account for the necessary permutation invariance of the yield function in the principal stress. 

We proceed by enforcing this constraint through permutation-invariant neural networks proposed in \cref{kimura2024permutation}, based on the Deep Set \cite{zaheer2017deep} framework. Recently, these permutation-invariant networks were used to formulate polyconvex strain energy functions in terms of principal stretches \cite{TEPOLE2025113469}. 
Here, we call this representation Permutation-Invariant - Input Convex Neural Networks (PI-ICNNs), which adopt the same architecture as in \cref{TEPOLE2025113469}. We make a distinction between two different methodologies adopted when using PI-ICNNs, which we denote by $\text{PI-ICNN}_1$ and $\text{PI-ICNN}_2$, respectively. The subscript here denotes the number of linear transformations employed for modeling the yield surface. Like Yld2004-18p, $\text{PI-ICNN}_2$ uses two linear transformations $\bm{L}'$ and $\bm{L}''$ that have the same form as Eq. \eqref{eq:projection_L} but independent components $c'_{ij}$ and $c''_{kl}$ respectively, whereas $\text{PI-ICNN}_1$ uses a single linear transformation $\bm{L}'$ with coefficients $c'_{mn}$. As we will show later, owing to the representative prowess of PI-ICNNs, we can obtain comparable, if not better, results with a single linear transformation of stresses as compared to Yld-200418p, which employs two such linear transformations. 

Assuming the network takes a vector $\bm{s}'$ of transformed principal deviatoric stresses as an input $[{s}_{1}', {s}_{2}', {s}_{3}']$, we can define $\text{PI-ICNN}_1$ as
\begin{equation} \label{Eq:PI-ICNN1}
    h({s}_{1}', {s}_{2}', {s}_{3}') = \mathcal{N}_{MC} \left(\left[\frac{1}{3} \sum_{i=1}^3  \mathcal{N}_{C} ({s}'_i)^p \right]^\frac{1}{p}  \right) ,
\end{equation}

where $\mathcal{N}_C$ is a simple ICNN and $\mathcal{N}_{MC}$ is an ICNN that is further restricted to be monotonically non-decreasing. The latter constraint can be enforced by ensuring that $\mathbf{W}_{0:k-1}^{[xx]} \geq 0$ in Eq. \eqref{eq:ICNN}. This essentially extends the constraints of a standard ICNN to include non-negativity of the weights in the first layer. Both $\mathcal{N}_C$ and $\mathcal{N}_{MC}$ are functions of scalar-valued inputs and give scalar-valued outputs. Lastly, the representation is completed by the trainable parameter $p$. 

Similarly, we can design the $\text{PI-ICNN}_2$ network to take as input the components of two vectors $\bm{s}'$ and $\bm{s}''$ obtained via two linear transformations (with independent unknown parameters) following Eq. \eqref{eq:Lineartransformations}. Then the network takes the form
\begin{equation} \label{Eq:PI-ICNN2}
    h(\bm{s}', \bm{s}'') = \mathcal{N}_{MC} \left(\left[\frac{1}{3} \sum_{i=1}^3  \bar{\mathcal{N}}_{C} ({s}'_i)^p \right]^\frac{1}{p} + \left[\frac{1}{3} \sum_{j=1}^3  \bar{\bar{\mathcal{N}}}_{C} ({s}''_j)^q \right]^\frac{1}{q}  \right) .
\end{equation}
with trainable parameters $p$ and $q$.
Again, we satisfy permutation invariance with respect to the components of each of these transformed stress vectors. The bars on top of $\mathcal{N}_{C}$ networks signify the use of different networks. Additionally, we have separate trainable parameters $p$ and $q$ for each network. Some other variants of $\text{PI-ICNN}_2$ that we have tested are detailed in Appendix \ref{app:PIICNN2_variants}.

\section{Dataset and calibration} \label{Sec:Dataset}
\begin{figure}
    \centering
    \includegraphics[width=1.0\linewidth]{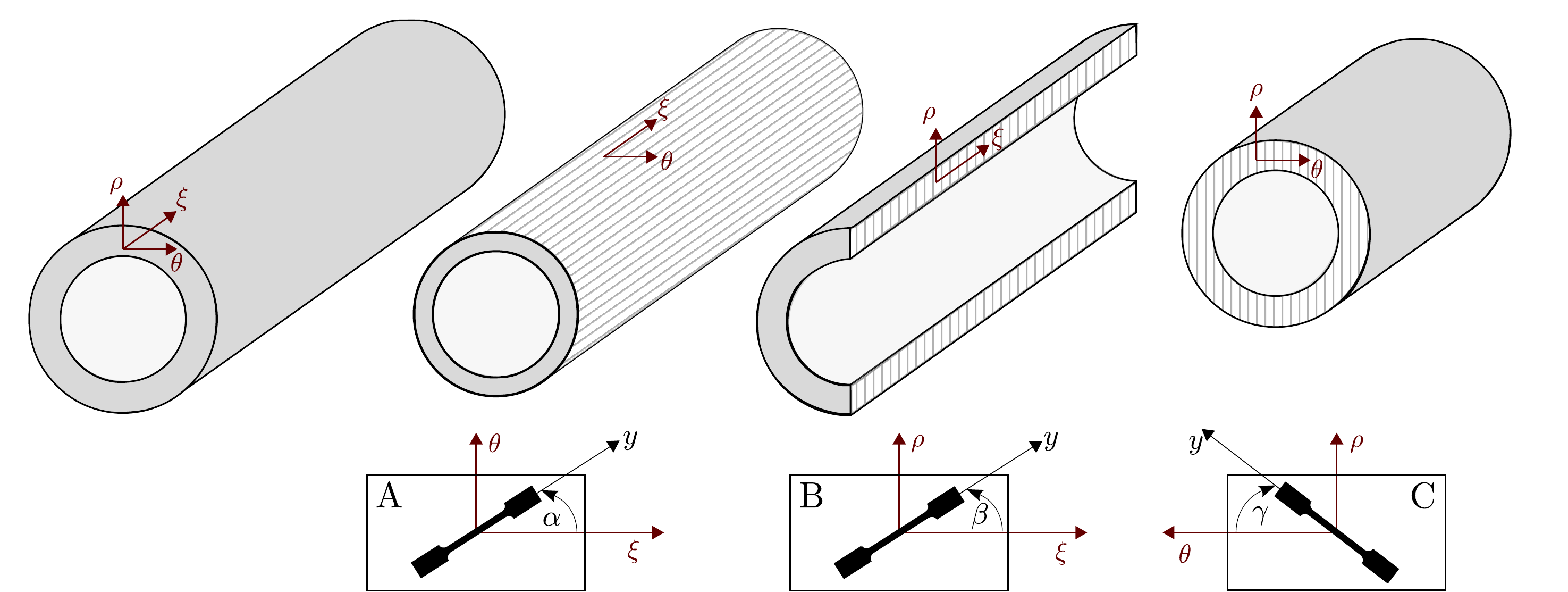}
    \caption{Sketch of the cylindrical extrusion showing the $\{\rho, \theta, \xi\}$ axes as well as the
orientations of the planes A, B, and C and the specimen orientation angles $\alpha$, $\beta$, and $\gamma$.}
    \label{fig:exp_setup}
\end{figure}

We calibrate our models on the 7079 aluminum dataset reported by Corona et.~al \cite{corona2021anisotropic}. Their study machined uniaxial tension specimens in twelve different orientations cut from an extruded Al-7079 tube, shown in Figure \ref{fig:exp_setup} and provided the yield values and Lankford coefficients for each of these orientations. Interested readers are referred to \cref{corona2021anisotropic} for a detailed overview of the experimental setup. Here, we summarize the process. As shown in Figure \ref{fig:exp_setup}, the axes $\{\rho, \theta, \xi\}$ are aligned with the through-thickness, circumferential, and axial directions of the extrusion. The axes $\{x,y,z\}$ are aligned with the specimen's geometry. The $y$ axis is defined along the longitudinal direction of the specimen. Three planes exist in the extrusion. We denote the $\xi$-$\theta$ plane as A, the $\xi$-$\rho$ plane as B, and the $\theta$-$\rho$ plane as C. The specimens were machined from these planes in different orientations controlled by the angles $\alpha$, $\beta$, and $\gamma$. Uniaxial tensile tests were conducted along the $y$ axis of each specimen to measure the corresponding yield stresses and Lankford coefficients.

In Table \ref{tab:corona_specimens}, we report the values obtained by Corona et.~al \cite{corona2021anisotropic}. We use $\sigma^c$ to denote the axial yield stress in the specimens and $r^{c}$ to denote the corresponding Lankford coefficients.

\begin{table}[h!]
\centering
\caption{Calibration data for the twelve Al 7079 specimens used in this study (reported in \cref{corona2021anisotropic}).}
\label{tab:corona_specimens}
\begin{tabular}{ccclc}
\toprule
Specimen & Plane & Orientation & $\sigma^{c}$ (MPa) & $r^{c}$ \\
\midrule
1  & A & $\alpha=0^{\circ}$ & 525($=Y_0$) & 0.18 \\
2  & A & $\alpha=15^{\circ}$  & 512 & 0.27 \\
3  & A & $\alpha=30^{\circ}$ &  515 & 0.75 \\
4  & A & $\alpha=45^{\circ}$&   505 & 1.20 \\
5  & A & $\alpha=60^{\circ}$&   493 & 1.00 \\
6  & A & $\alpha=75^{\circ}$&   511 & 0.70 \\
7  & A & $\alpha=90^{\circ}$&   530 & 0.91 \\
8  & B & $\beta=45^{\circ}$&   510 & 2.90 \\
9  & B & $\beta=60^{\circ}$&   544 & 1.50 \\
10 & B & $\beta=90^{\circ}$&   523 & 1.10 \\
11 & C & $\gamma=45^{\circ}$&   486 & 0.47 \\
12 & C & $\gamma=60^{\circ}$&   485 & 0.52 \\
\bottomrule
\end{tabular}
\end{table}

We adopt a slightly adjusted loss function $\mathcal{L}$ to the one reported by Corona et.~al, motivated by previous studies \cite{BARLAT20051009, tardif2012determination} which is of the form
\begin{equation}
    \mathcal{L} = \mathcal{L}_{\sigma} + \mathcal{L}_r \ ,
\end{equation}
where 
\begin{equation}
    \mathcal{L}_{\sigma} = \sum_i^N w_{\sigma} \left( \frac{\hat{\Phi}(\sigma^c_i)}{Y} - 1  \right)^2 = \sum_i^N w_{\sigma} [\hat{f}(\sigma^c_i)]^2 \ ,
\end{equation}
accounts for the error in the yield stress and 
\begin{equation}
    \mathcal{L}_r = \sum_i^N w_{r} \left( \frac{\hat{r}(\sigma^c_i)}{r^c_i} - 1  \right)^2 \ ,
\end{equation}
represents a weighted error in the Lankford coefficients. 

The superscript $\hat{\bullet}$ is used to denote predicted values (consider Eq. \eqref{eq:Yield_general} without hardening). The scalars $w_{\sigma}$ and $w_{r}$ are used to weight the contributions of each loss component. We follow Refs. \cite{BARLAT20051009,corona2021anisotropic} and fix $w_{\sigma}=10$ and $w_r=1$. 
We remark that we show formulations of $ \mathcal{L}_{\sigma}$ both in terms of the equivalent stress function $\Phi(\bm{\sigma})$ and the yield function $f(\bm{\sigma})$. This is necessary because we calibrate the equivalent stress function for the phenomenological models and the hybrid approach. On the other hand, we directly model the (entire) yield function for the neural network approaches, i.e., ICNNs and PI-ICNNs.

The total number of trainable parameters for each method is reported in Table \ref{tab:no_parameters}. We see an additional trainable parameter for Yld2004-18p because we also treat the exponent $a$ in Eq. \eqref{eq:Yld2004-18p} as a trainable parameter. 
The ICNN and the hybrid approach differ only by six trainable parameters, with the additional parameters being the coefficients needed to calibrate the base model Hill-48. We also highlight that nine additional linear transformation parameters are needed for $\text{PI-ICNN}_1$ and 18 for $\text{PI-ICNN}_2$.

Lastly, we note that the NN-based models all have a similar number of trainable parameters. Thereby, we hope to offer a fair comparison. 
 A separate study on the influence of the number of trainable parameters is presented in Appendix \ref{app:paramStudy}.

\begin{table}[h!]
\centering
\caption{Number of trainable parameters for each approach.}
\label{tab:no_parameters}
\begin{tabular}{lc}
\toprule
Method & Number of trainable parameters \\
\midrule
Hill-48         & 6  \\
Yld2004-18p       & 19 \\
ICNN         & 257  \\
Hybrid       & 263   \\
$\text{PI-ICNN}_1$    & 234  \\
$\text{PI-ICNN}_2$    & 272  \\
\bottomrule
\end{tabular}
\end{table}

 For all methods that involve neural networks, we choose the Adam optimizer \cite{kingma2014adam}.
 A constant learning rate of 5$\times 10^{-3}$ is employed throughout. 
However, in terms of fitting coefficients of phenomenological models, (first-order) gradient-based methods have been observed to be susceptible to getting stuck at local minima and to be heavily dependent on initial values \cite{chaparro2008material, hariharan2014novel,fuhg2023modular}. We observed similar trends when trying to fit Hill-48 and Yld2004-18p with Adam as well as a quasi-Newton optimizer (L-BFGS \cite{byrd1995limited}). Therefore, we instead rely on a Genetic Algorithm to ensure robust optimization of the coefficients of Hill-48 and
Yld2004-18p. In particular, we use the Covariance Matrix Adaptation Evolution Strategy (CMA-ES), see Refs. \cite{cmaes, Hansen16a} for more information.
We note that the anisotropy coefficients are not unique (see, e.g., \cref{zhang2022parameter}). This means combinations of different values can converge to the same yield stress and Lankford coefficients. 
We do not pursue a calibration strategy in this work that enforces uniqueness.

\begin{table}[h!]
\centering
\caption{Average error metrics for CMA-ES.}
\label{tab:cmaes_corona}
\begin{tabular}{lcccc}
\toprule
Method & Max $|f|$ & Mean $|f|$ & Max $|\Delta r|$ & Mean $|\Delta r|$ \\
\midrule
Hill-48 (Corona et.~al)         & 87.24 & 39.67 & 0.89 & 0.22 \\
Hill-48 (CMA-ES)        & 86.57 & 39.53 & 0.90 & 0.22 \\
Yld2004-18p (Corona et.~al)      & 18.43 & 7.47  & 0.19 & 0.06 \\
Yld2004-18p (CMA-ES)      & 14.93 & 6.36  & 0.16 & 0.04 \\
\bottomrule
\end{tabular}
\end{table}

We verify our methodology by comparing the optimization results for Hill-48 and Yld-200418p for the entire dataset against the yield and Lankford coefficients reported in \cref{corona2021anisotropic}. The results are shown in Table \ref{tab:cmaes_corona}.
We present the mean values obtained across three different parameter initializations for CMA-ES. 

It can be seen that there is a small discrepancy in the maximum and mean yield and the Lankford coefficients when compared with those reported by Corona et al. We use the entire dataset here to make the comparison since the same was done by Corona. et.~al. However, for further evaluation of the proposed methodologies and their comparisons with phenomenological models, the dataset will be split into training, validation, and test datasets. In particular, nine random splits are made to generate nine datasets as reported in Table \ref{tab:dataset_splits}, with eight specimens in each training dataset, and two in each validation and test dataset. Since we have overparameterized regression models, we employ a slightly modified form of a classical early stopping strategy based on the validation loss \cite{prechelt2002early}. In particular,
\textit{generalization} is quantified as the percentage increase in validation loss at epoch $i$ compared to the lowest validation loss recorded up to that point in training.
Formally, we can write it as:
\begin{equation}
    \mathcal{L}_{\text{gen}}(i) = 100 \cdot \left( \frac{\mathcal{L}_{\text{val}}(i)}{\text{min} (\mathcal{L}_{\text{val}}(1:i))} - 1 \right) \ .
\end{equation}
This allows us to define the generalization criterion by checking if $\mathcal{L}_{\text{gen}}(i)$ exceeds a threshold value $\alpha$, i.e., we choose the epoch $i$ for which
\begin{equation}
    \mathcal{L}_{\text{gen}}(i) > \alpha \ .
    \label{eq:earlyStopCriteria}
\end{equation}
We remark that classical stopping based on an increase in validation loss can be obtained by setting $\alpha=0$. Note that other techniques, e.g., sparsification via regularization \cite{fuhg2024extreme, flaschel2021unsupervised}, can also be employed to counter overfitting.

Early stopping strategies are based on the presumption of high training times. Early stopping then helps avoid overfitting, while at the same time reducing long training times. However, our networks are relatively small and are not as computationally expensive to train. Therefore, we train the networks for a high number of epochs ($N=50000$).
We therefore have access to the validation loss over the full training duration and can retroactively pick out a model at a certain point during training.
We then choose the highest epoch that satisfies Eq. \eqref{eq:earlyStopCriteria}. Since we are not constrained by high training times, this allows us to potentially choose a better model in case validation loss increases, and then decrease again. 

In this work, we report results for $\alpha=0$, which gives us the epoch with the lowest validation loss, as well as for $\alpha=10$. This strategy is only applied for the NN-based approaches since the phenomenological models do not exhibit overfitting, as can be seen in Appendix \ref{app:loss_evolution}, but contain a lot of noise in the validation loss, especially at the earlier epochs, due to the use of the evolutionary strategy.

\begin{table}[h!]
\centering
\caption{Specimens used for training, validation, and test in the nine dataset splits where specimen numbering follows Table~\ref{tab:corona_specimens}.}
\label{tab:dataset_splits}
\begin{tabular}{lccc}
\toprule
Dataset & Training specimens & Validation specimens & Test specimens \\  
\midrule
1  & 1,\,2,\,3,\,5,\,6,\,7,\,9,\,12   & 4,\,8  & 10,\,11 \\
2  & 1,\,2,\,4,\,6,\,7,\,8,\,10,\,11  & 9,\,12 & 3,\,5   \\
3  & 1,\,2,\,4,\,7,\,8,\,10,\,11,\,12 & 5,\,9  & 3,\,6   \\
4  & 1,\,3,\,4,\,5,\,7,\,8,\,9,\,11   & 2,\,10 & 6,\,12  \\
5  & 2,\,4,\,5,\,6,\,7,\,8,\,9,\,12   & 1,\,11 & 3,\,10  \\
6  & 1,\,2,\,5,\,6,\,8,\,9,\,10,\,12  & 3,\,7  & 4,\,11  \\
7  & 2,\,3,\,4,\,6,\,7,\,9,\,11,\,12  & 1,\,10 & 5,\,8   \\
8  & 1,\,2,\,4,\,5,\,7,\,9,\,10,\,11  & 3,\,6  & 8,\,12  \\
9 & 3,\,4,\,5,\,6,\,8,\,9,\,10,\,12  & 1,\,7  & 2,\,11  \\
\bottomrule
\end{tabular}
\end{table}

\section{Results} \label{Sec:Results}

To keep the results concise, we report the averages of the error metrics across all nine datasets laid out in the previous section. This includes the maximum and mean discrepancies of the yield function and the r-values. The latter is denoted with $\Delta r$. Training, validation, and test set averages for $\alpha=10$ are reported in Tables \ref{tab:training_lowerParam_alpha10}, \ref{tab:validation_lowerParam_alpha10}, and \ref{tab:test_lowerParam_alpha10}, respectively. Results with $\alpha=0$, i.e., the epoch with the lowest validation loss, are reported in Appendix \ref{app:alphaequal0}. For the training dataset, Hill-48 performs the worst in terms of fitting both the yield and the r-values. ICNN and the Hybrid approach yield comparable results. Barlat's Yld2004-18p function has the lowest training loss, followed by the PI-ICNN-based frameworks. The error metrics for the entire dataset (without any splits) are reported in Appendix \ref{app:completeDataset}. In that case, ICNN and the hybrid approach match both the yield values and the Lankford coefficients almost perfectly. This indicates that both approaches are prone to over-fitting, which is demonstrated by their poor performance on the validation and test sets as reported in Tables \ref{tab:validation_lowerParam_alpha10} and \ref{tab:test_lowerParam_alpha10}. Although they outperform Hill-48 on the training dataset, Hill-48 only has six trainable parameters and therefore shows significantly better results on the validation and test sets. However, it still falls short of the accuracy demonstrated by Yld2004-18p and the PI-ICNN approaches. 

\begin{figure}
    \centering
    \includegraphics[width=1.0\linewidth]{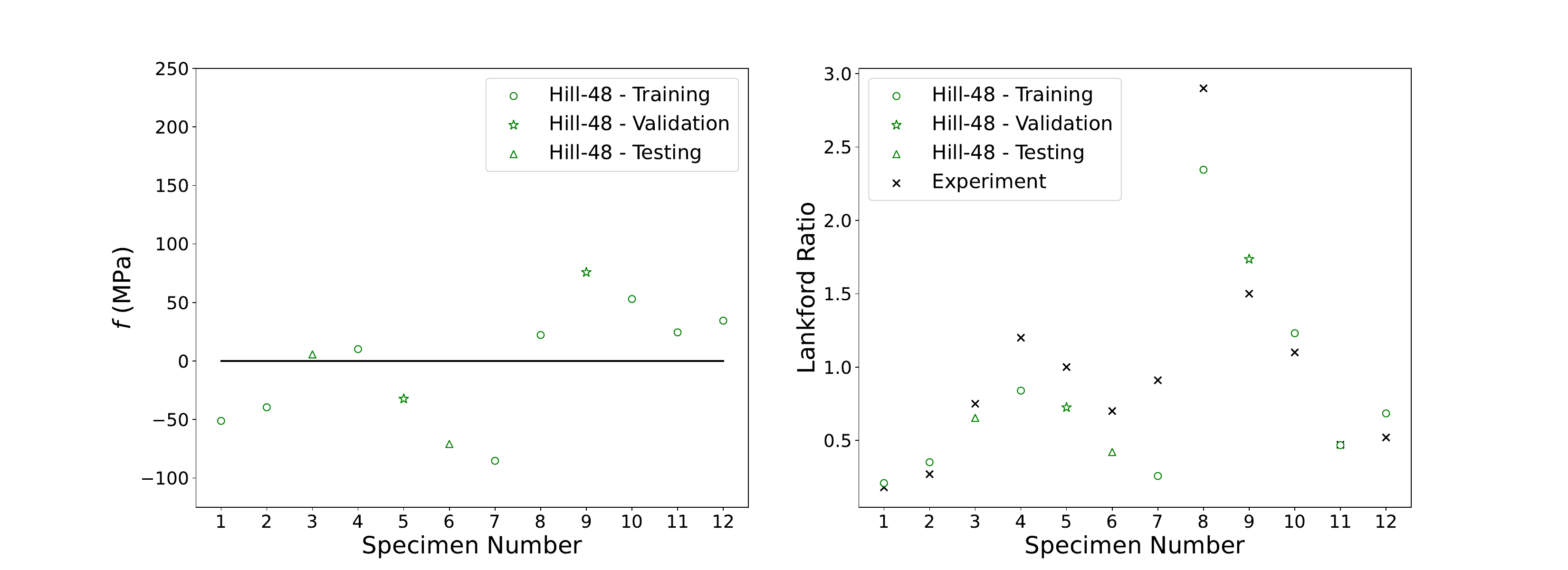}
    \caption{Comparison of Hill-48 predictions with experimental measurements for dataset 3.}
    \label{fig:YieldandLr_hill}
\end{figure}

\begin{figure}
    \centering
    \includegraphics[width=1.0\linewidth]{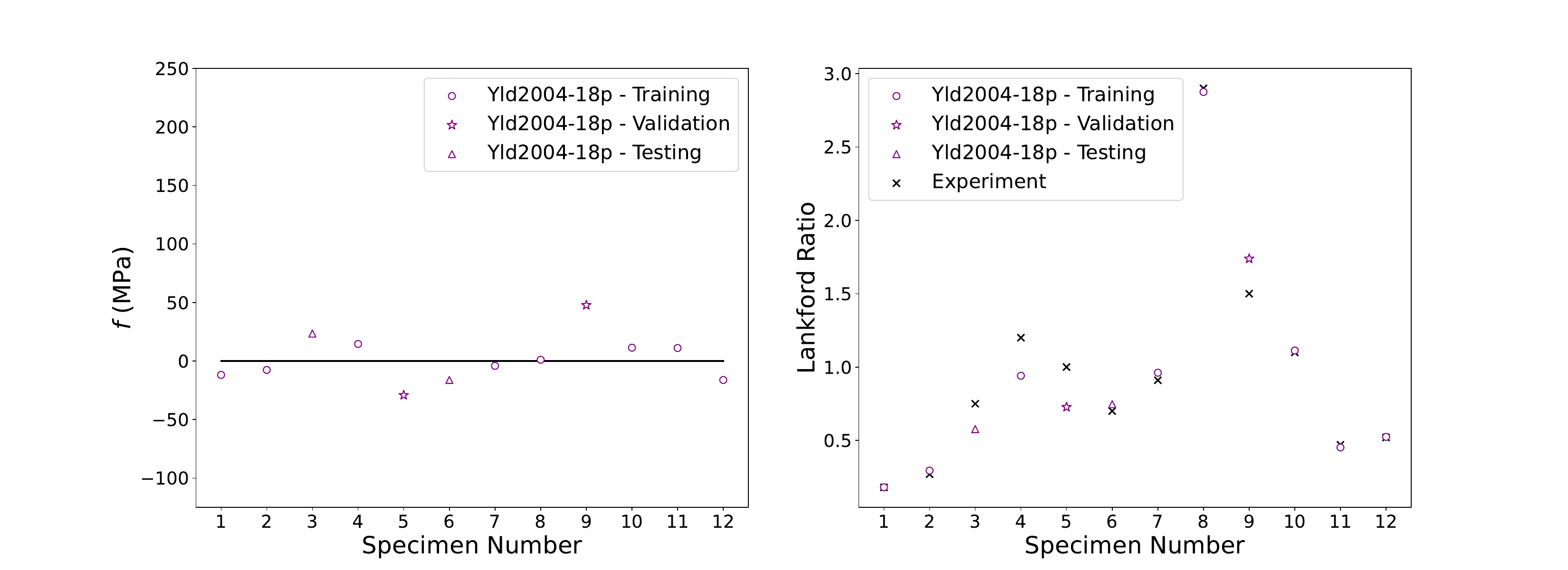}
    \caption{Comparison of Yld2004-18p predictions with experimental measurements for dataset 3.}
    \label{fig:YieldandLr_barlat}
\end{figure}

\begin{figure}
    \centering
    \includegraphics[width=1.0\linewidth]{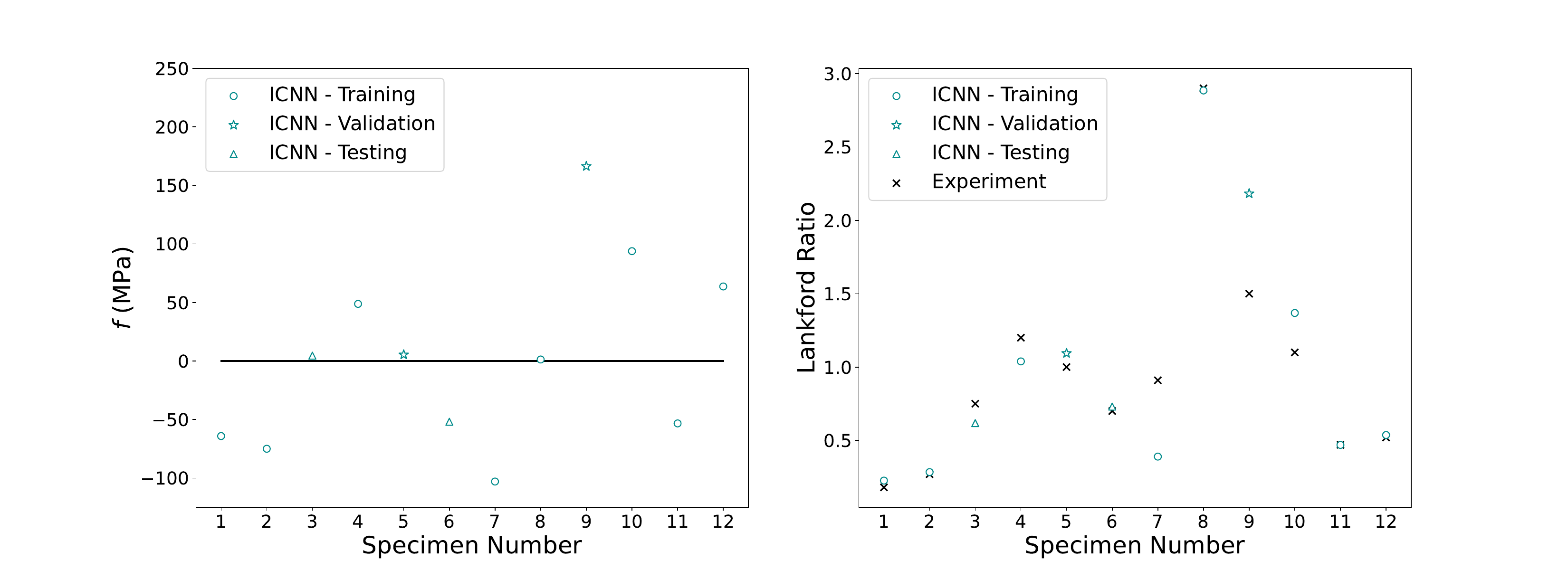}
    \caption{Comparison of pure ICNN-based framework predictions with experimental measurements for dataset 3.}
    \label{fig:YieldandLr_icnn}
\end{figure}

\begin{figure}
    \centering
    \includegraphics[width=1.0\linewidth]{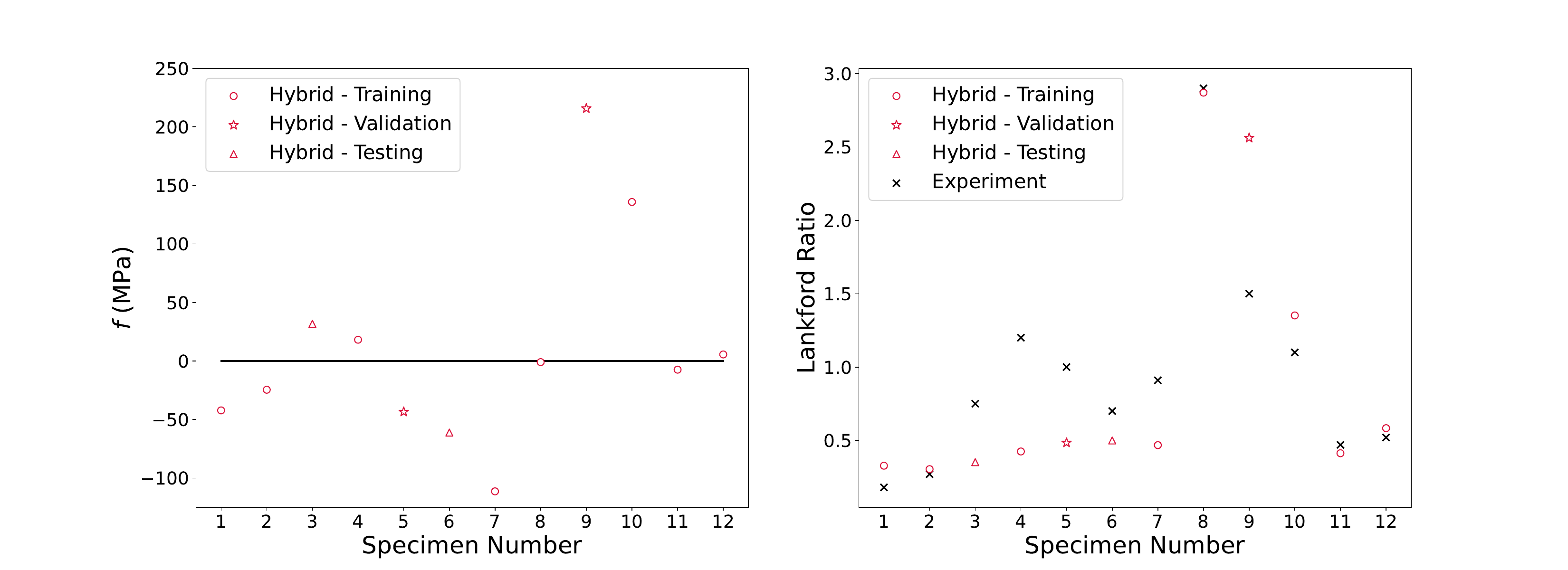}
    \caption{Comparison of the hybrid approach predictions with experimental measurements for dataset 3.}
    \label{fig:YieldandLr_hybrid}
\end{figure}

\begin{figure}
    \centering
    \includegraphics[width=1.0\linewidth]{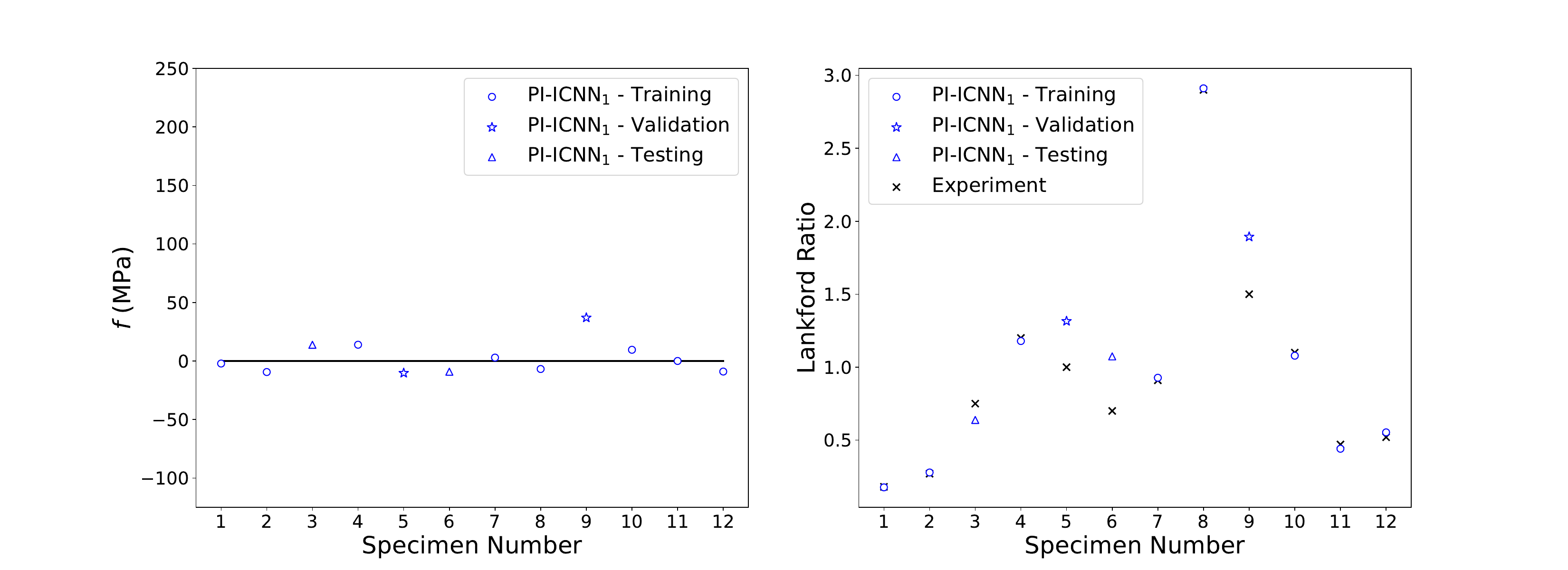}
    \caption{Comparison of $\text{PI-ICNN}_1$ predictions with experimental measurements for dataset 3.}
    \label{fig:YieldandLr_piicnn1}
\end{figure}

\begin{figure}
    \centering
    \includegraphics[width=1.0\linewidth]{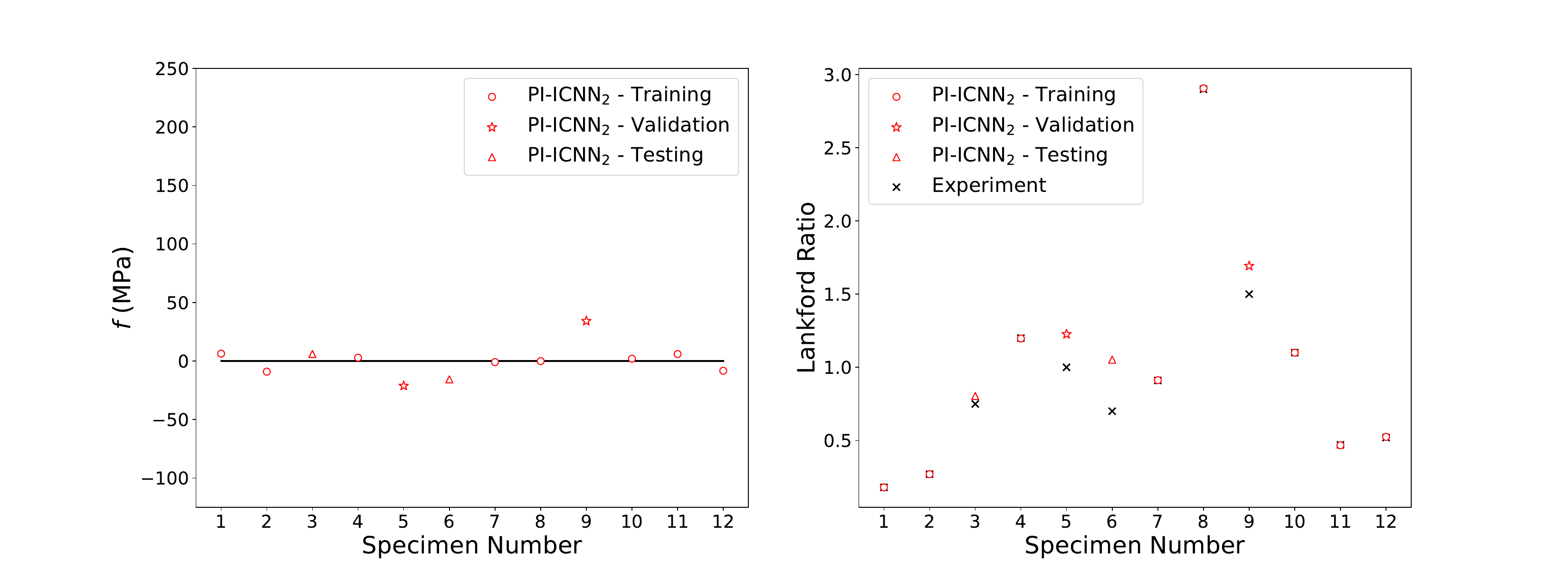}
    \caption{Comparison of $\text{PI-ICNN}_2$ predictions with experimental measurements for dataset 3.}
    \label{fig:YieldandLr_piicnn2}
\end{figure}

We can report that the \text{PI-ICNN} frameworks seem to outperform Yld2004-18p in the validation and test set.  To visualize the performance of all these approaches, we show the predicted yield along with the true and predicted r-values for one randomly chosen dataset split, specifically dataset 3, in Figures \ref{fig:YieldandLr_hill}-\ref{fig:YieldandLr_piicnn2}. The evolution of loss and stoppage epochs for the same dataset, with $\alpha=0$ and $\alpha=10$ in Eq. \eqref{eq:earlyStopCriteria}, is also reported in Appendix \ref{app:loss_evolution}. The performance of each framework on the test datasets in all nine splits is documented in Appendix \ref{testDataset_Figures}. 

\begin{table}[h!]
\centering
\caption{Error metrics for the training dataset with $\alpha=10$ in Eq. \eqref{eq:earlyStopCriteria}}.
\label{tab:training_lowerParam_alpha10}
\begin{tabular}{lcccc}
\toprule
Method & Max $|f|$ & Mean $|f|$ & Max $|\Delta r|$ & Mean $|\Delta r|$ \\
\midrule
Hill-48         & 90.14 & 42.25 & 0.99 & 0.29 \\
Yld2004-18p       & \textbf{13.23} & \textbf{5.55}  & 0.25 & 0.04 \\
ICNN         & 51.78 & 22.43 & 0.24 & 0.07 \\
Hybrid       & 42.98  & 13.50  & 0.30 & 0.09 \\
$\text{PI-ICNN}_1$    & 32.46  & 14.34  & 0.17 & 0.05 \\
$\text{PI-ICNN}_2$    & 19.74  & 7.44   & \textbf{0.10} & \textbf{0.03} \\
\bottomrule
\end{tabular}
\end{table}

\begin{table}[h!]
\centering
\caption{Error metrics for the validation dataset with $\alpha=10$ in Eq. \eqref{eq:earlyStopCriteria}}.
\label{tab:validation_lowerParam_alpha10}
\begin{tabular}{lcccc}
\toprule
Method & Max $|f|$ & Mean $|f|$ & Max $|\Delta r|$ & Mean $|\Delta r|$ \\
\midrule
Hill-48         & 71.95 & 51.87 & 0.46 & 0.31 \\
Yld2004-18p       & 48.58  & 31.19  & 2.51 & 1.34 \\
ICNN         & 138.64  & 92.78  & 0.52 & 0.32 \\
Hybrid       & 184.6  & 132.11  & 0.77 & 0.50 \\
$\text{PI-ICNN}_1$    & \textbf{23.02}  & \textbf{16.97}  & \textbf{0.32} & \textbf{0.21} \\
$\text{PI-ICNN}_2$    & 32.92 & 23.55 & 0.35 & \textbf{0.21} \\
\bottomrule
\end{tabular}
\end{table}

\begin{table}[h!]
\centering
\caption{Error metrics for the test dataset with $\alpha=10$ in Eq. \eqref{eq:earlyStopCriteria}}.
\label{tab:test_lowerParam_alpha10}
\begin{tabular}{lcccc}
\toprule
Method & Max $|f|$ & Mean $|f|$ & Max $|\Delta r|$ & Mean $|\Delta r|$ \\
\midrule
Hill-48         & 57.09 & 45.88 & 1.50 & 0.91 \\
Yld2004-18p       & 48.03 & 33.32 & 1.04 & 0.59 \\
ICNN         & 168.27  & 112.10  & 1.14 & 0.68 \\
Hybrid       & 210.88  & 142.43  & 1.00 & 0.64 \\
$\text{PI-ICNN}_1$    & \textbf{33.00}  & \textbf{23.73}  & \textbf{0.31} & \textbf{0.18} \\
$\text{PI-ICNN}_2$    & 37.71 & 23.97 & 0.64 & 0.35 \\
\bottomrule
\end{tabular}
\end{table}

\begin{figure}[htbp]
  \centering
  \begin{subfigure}{0.48\textwidth}
    \centering
    \includegraphics[width=\linewidth]{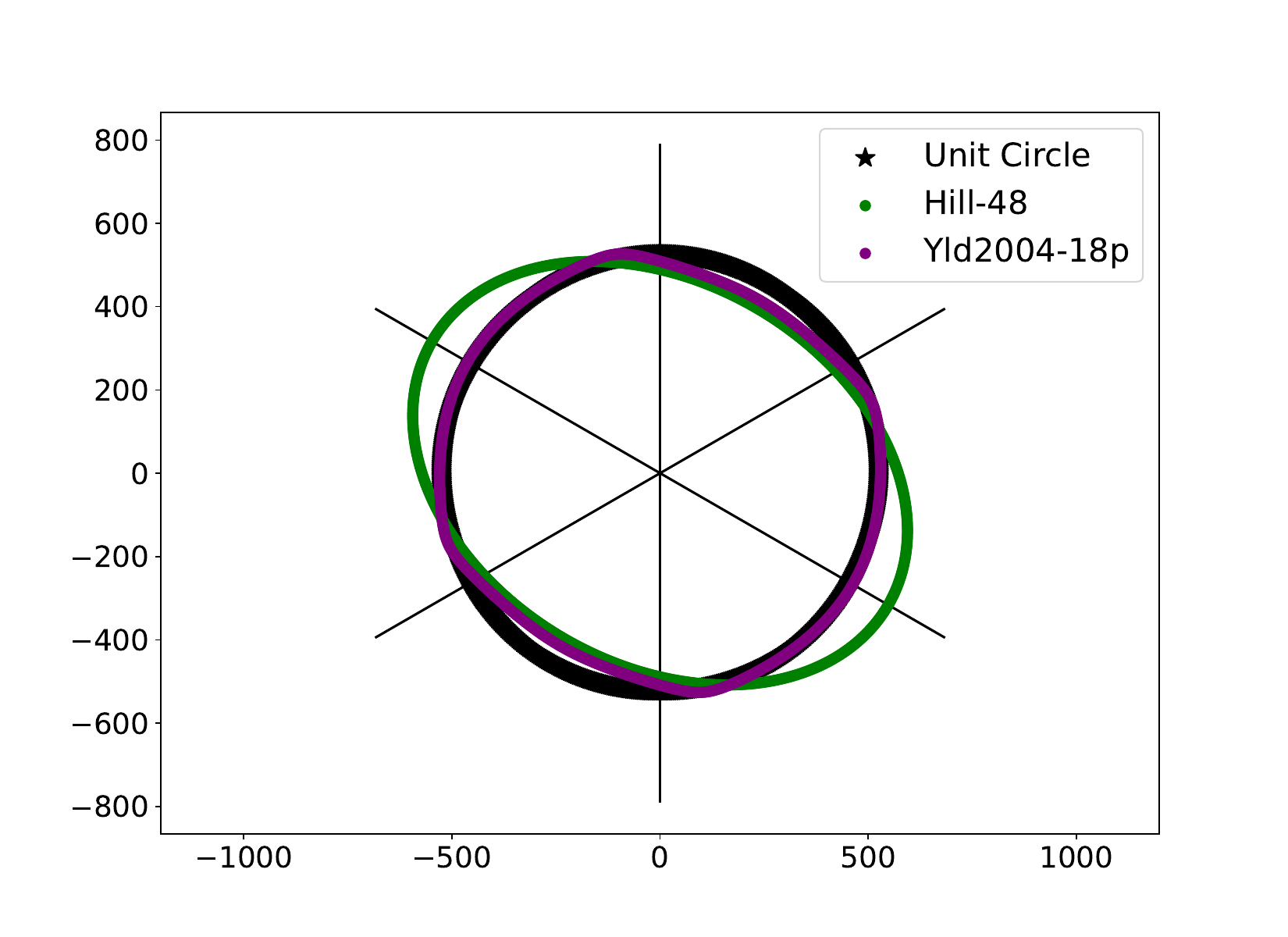}
    \caption{}
    \label{fig:PiPlane_plots_phen_spec8}
  \end{subfigure}\hfill
  \begin{subfigure}{0.48\textwidth}
    \centering
    \includegraphics[width=\linewidth]{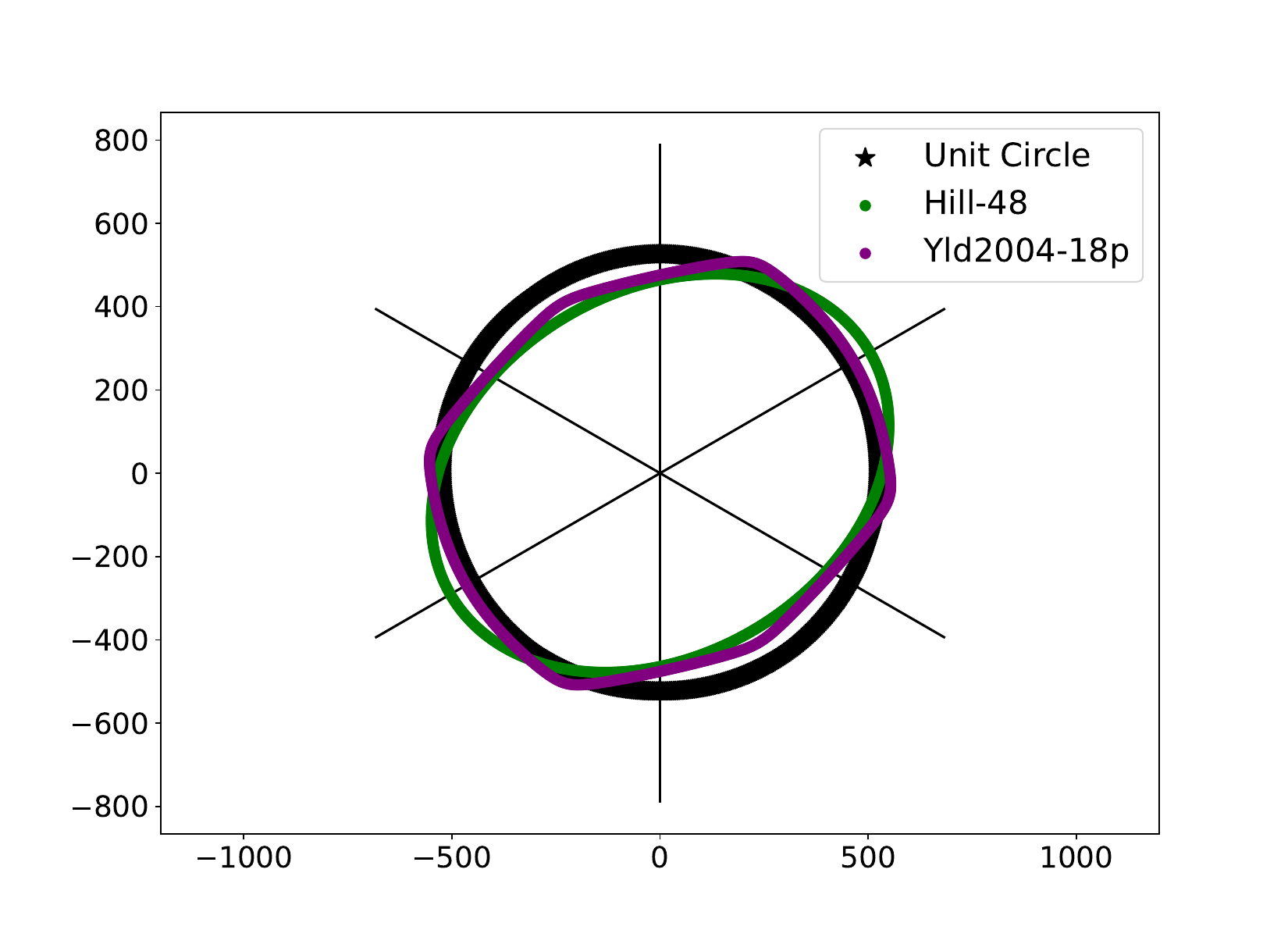}
    \caption{}
    \label{fig:PiPlane_plots_phen_spec11}
  \end{subfigure}

  \vspace{1ex}  

  \begin{subfigure}{0.48\textwidth}
    \centering
    \includegraphics[width=\linewidth]{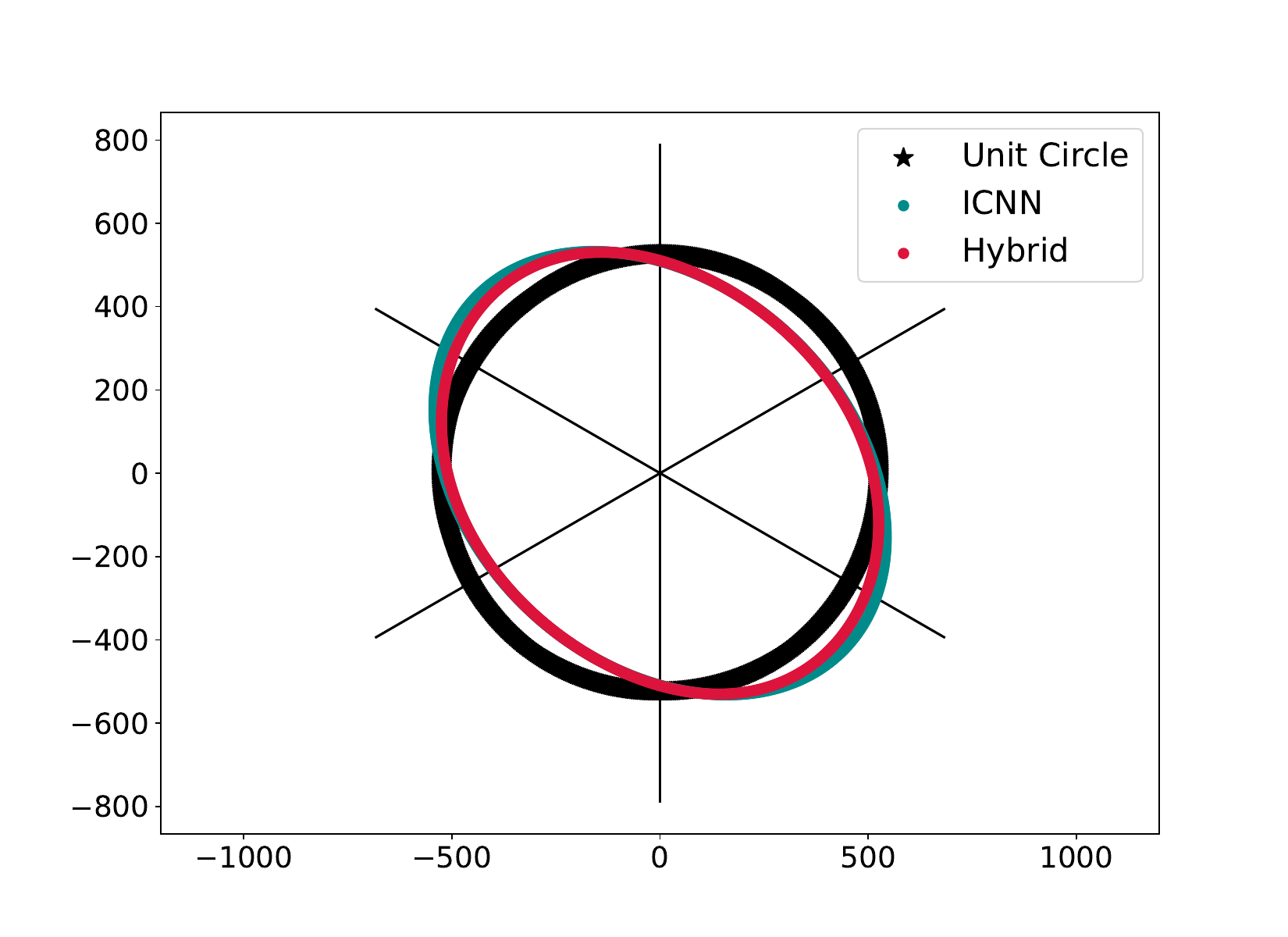}
    \caption{}
    \label{fig:PiPlane_plots_icnnhybrid_spec8}
  \end{subfigure}\hfill
  \begin{subfigure}{0.48\textwidth}
    \centering
    \includegraphics[width=\linewidth]{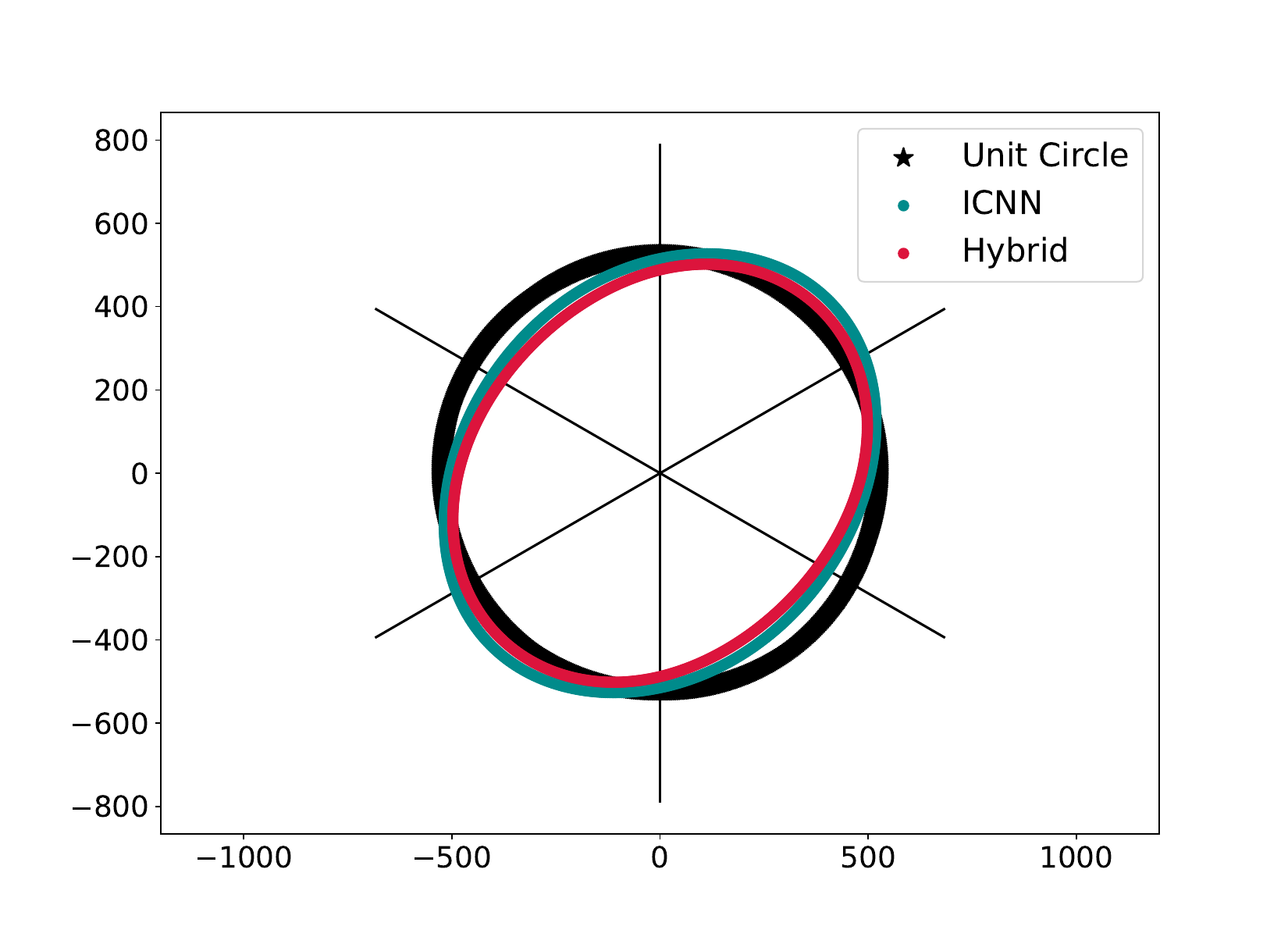}
    \caption{}
    \label{fig:PiPlane_plots_icnnhybrid_spec11}
  \end{subfigure}

  \vspace{1ex}

  \begin{subfigure}{0.48\textwidth}
    \centering
    \includegraphics[width=\linewidth]{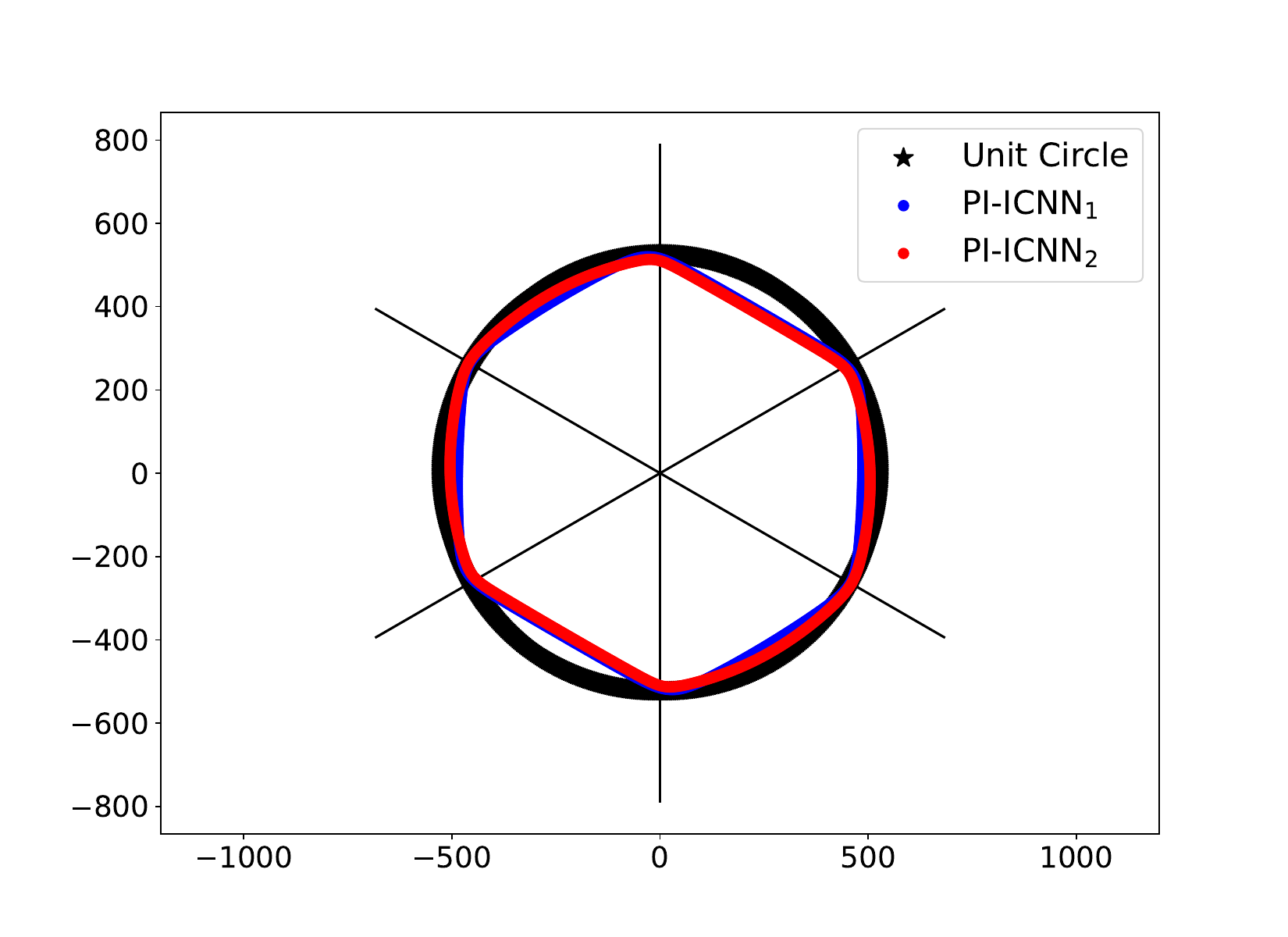}
    \caption{}
    \label{fig:PiPlane_plots_picnn_spec8}
  \end{subfigure}\hfill
  \begin{subfigure}{0.48\textwidth}
    \centering
    \includegraphics[width=\linewidth]{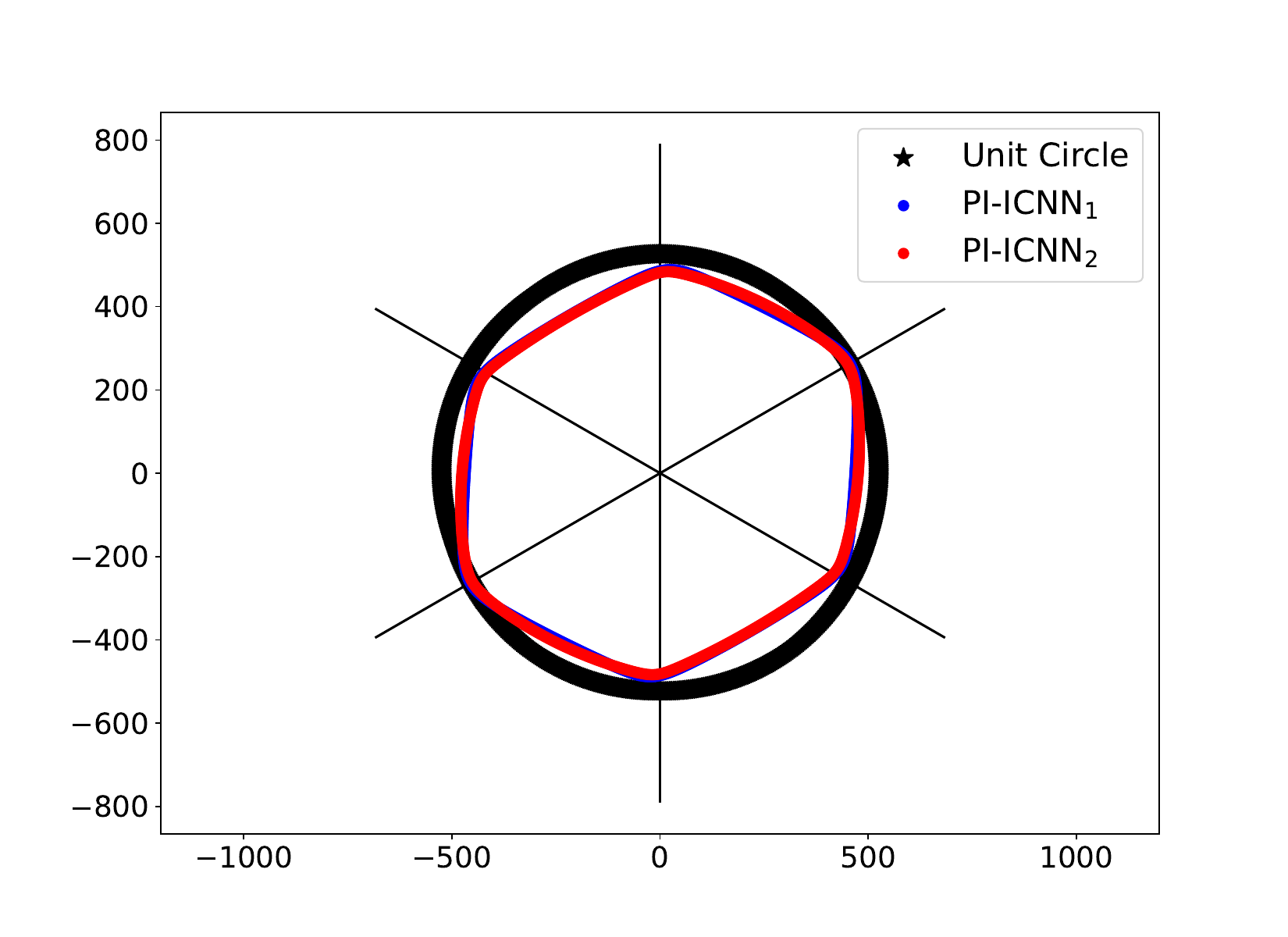}
    \caption{}
    \label{fig:PiPlane_plots_picnn_spec11}
  \end{subfigure}

  \caption{Deviatoric plane plots for (a,c,e) specimen 8 and (b,d,f) specimen 11 in dataset 3.}
  \label{fig:PiPlane_plots_complete}
\end{figure}

Furthermore, in order to visually inspect the anisotropy of the modeled yield functions, along with other aspects such as convexity and smoothness, we plot these functions in the deviatoric plane. Details on how to obtain mappings to and from the deviatoric plane are outlined in Appendix \ref{section:dev_plane}. Here, again for the compactness of the results, we choose to show the deviatoric plane plots for only two specimens from dataset 3. Specifically, yield function plots are shown for specimens 8 and 11 since all frameworks decently predict the yield and r-values for these specimens. Due to the lack of ground truth data, this allows to make a comparison of the deviatoric plane plots obtained with various frameworks discussed earlier. Figures \ref{fig:PiPlane_plots_phen_spec8}, \ref{fig:PiPlane_plots_icnnhybrid_spec8} and \ref{fig:PiPlane_plots_picnn_spec8} show the deviatoric plane plots for specimen 8 whereas Figures \ref{fig:PiPlane_plots_phen_spec11}, \ref{fig:PiPlane_plots_icnnhybrid_spec11} and \ref{fig:PiPlane_plots_picnn_spec11} show these plots for specimen 11. From these figures, all the yield functions clearly appear to be convex and smooth. Moreover, while there exists a discrepancy in the shapes of the yield functions for different frameworks, they all display similar anisotropic tendencies, i.e., each surface resembles an ellipse with similarly oriented principal axes. This suggests that the directional dependence of yielding is preserved across all formulations.

\begin{figure}
    \centering
    \includegraphics[width=1.0\linewidth]{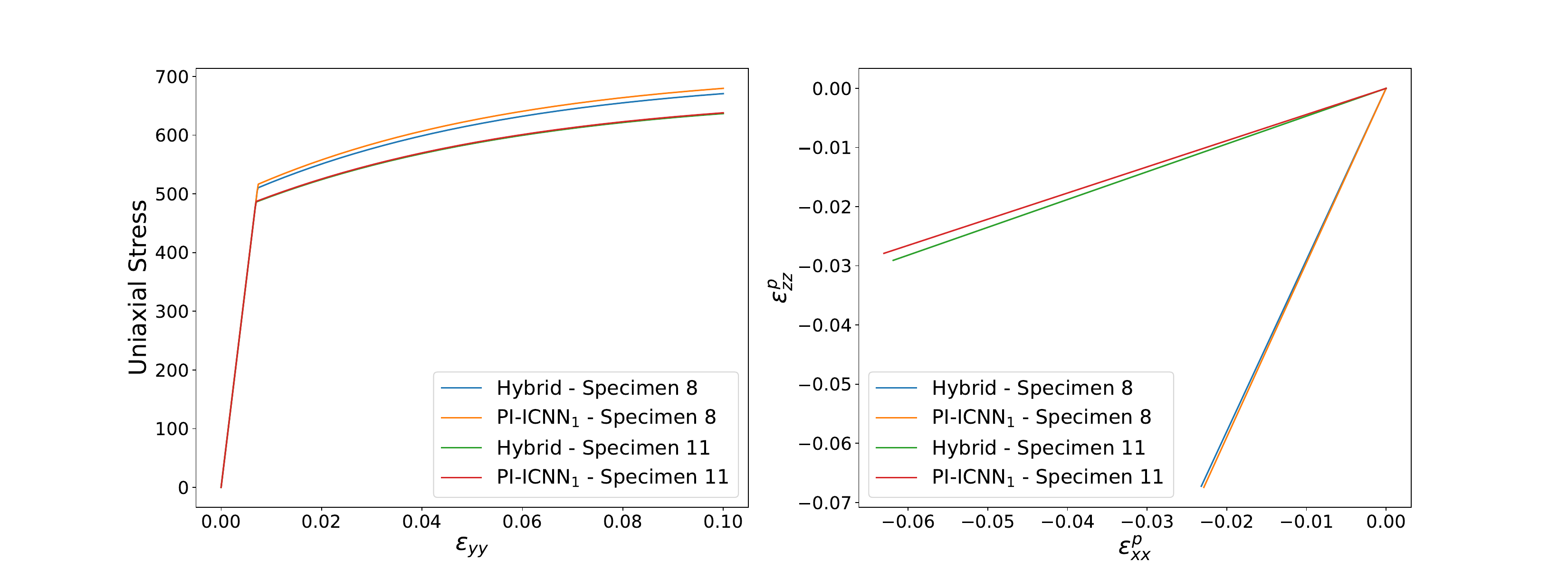}
    \caption{Uniaxial material-point simulations for specimens 8 and 11 from dataset 3 with the hybrid and $\text{PI-ICNN}_1$ yield functions. The figure shows the axial stress-strain response (left) as well the the evolution of off-axis plastic strains (right).}
    \label{fig:uniaxial_loading}
\end{figure}

Beyond the error metrics and visual inspections of the modeled yield functions, it is essential to demonstrate the practicality of the proposed approaches. This practicality is synonymous with their integration within a standard finite element (FE) framework, which employs a return mapping algorithm to ensure the consistency conditions laid out in Eq. \eqref{KKT}. To this end, we conducted uniaxial loading simulations in a standalone material point solver for the same two specimens as before, i.e., specimens 8 and 11 from dataset 3. While the full FE implementation would additionally require the consistent material tangent (which we do not consider here), the successful integration of our models into the return mapping procedure gives us confidence in the practical viability of our approaches. Since the goal here is only to demonstrate the applicability of the proposed methodologies and not to accurately predict the pre- or post-yield response, we calibrate neither the elastic constants nor the hardening behavior. Instead, we choose the same elastic constants as the ones reported by Corona et.~al, i.e., $E=70\text{GPa}$ and $\nu=0.33$, where $E$ and $\nu$ represent the Young's modulus and Poisson's ratio. Additionally, we employ a variant of Voce's hardening model \cite{kumar2025comparative}, which modifies the yield value as
\begin{equation}
    Y = Y_0 + R_{\infty} \left(1 - \exp \left( \gamma \bar\epsilon^p \right) \right) \ ,
\end{equation}
where $\bar\epsilon^p$ is the equivalent plastic strain, $Y_0$ denotes the initial yield, $R_{sat}$ represent the saturation stresses and $\gamma$ governs the rate of change of hardening. For the uniaxial loading simulation, we use $R_{sat}=200$ and $\gamma=20$. Finally, the framework laid out in Section \ref{Sec:Methods} is employed to ensure shape-consistent scaling of the yield surface. These simulations are presented in Figure \ref{fig:uniaxial_loading} for the hybrid and $\text{PI-ICNN}_1$ approaches. Since ICNNs are a part of the hybrid approach and both $\text{PI-ICNN-based}$ approaches only differ in the number of linear transformations, these simulations adequately show that all the proposed frameworks can be implemented in material point routines. Both approaches result in similar stress-strain response as evident from Figure \ref{fig:uniaxial_loading}. Additionally, the off-axis plastic strain plots also closely match, and the linearity present in these plots confirms that the yield function retains its original shape under isotropic hardening. 

\section{Conclusion and outlook}
In this work, we proposed novel frameworks for modeling thermodynamically consistent anisotropic yield functions and compared them with existing approaches. Physics-augmented neural network architectures, ICNNs and PI-ICNNs, were employed to get yield function representations in the six-dimensional stress space and the principal stress space, respectively. These architectures were designed to ensure thermodynamic consistency of the yield function representations. A correction (model-data-driven) approach was also investigated, which combined Hill-48 and ICNNs. Although this hybrid approach, much like the pure ICNN-based approach, fits the training data almost perfectly, it is prone to overfitting and therefore shows subpar performance on the validation and test datasets. Both approaches show similar behavior with sufficient network sizes. However, for networks with smaller sizes, the hybrid approach outperforms pure ICNNs on the training dataset owing to the additional information imparted to it through the Hill-48 model. Even though they are regularized by the physics-augmentation, overfitting encountered by these approaches can partly be attributed to the fact that both these approaches take input from the six-dimensional stress space. Therefore, when trained on only 8 data points, these networks only see a very small part of the input space. In contrast, Yld2004-18p, which assumes a phenomenological yield function form, takes input from the principal stress space and outperforms the aforementioned approaches on the investigated dataset. PI-ICNNs are similar to the Yld2004-18p in the sense that they not only share the same input space, but also model anisotropy into the yield function through the use of linear transformations. However, these networks do not assume any particular yield function form \textit{a priori}. They are trained to simultaneously learn an isotropic function and the linear transformations. This results in improved representative capabilities evident by the improved performance of $\text{PI-ICNN}_1$ and $\text{PI-ICNN}_2$  compared to Yld2004-18p. In particular, both approaches significantly outperform Yld2004-18p in fitting the Lankford coefficients.

Although the hybrid approach showed signs of overfitting, it still holds promise since ICNNs performed much worse on their own with a very small number of trainable parameters. The information provided by the underlying phenomenological model seems to constrain the ICNN during the initial part of training. This results in a lower initial loss and delayed overfitting. In contrast, the pure ICNN-based approach has relatively high initial errors and thereby starts overfitting earlier.

In future work, we intend to systematically study the nature of anisotropy in the yield functions by investigating the linear transformation itself. This could, for example, be used for designing materials with target anisotropies and to investigate the relationship between elastic and plastic anisotropies (see e.g. Refs. \cite{ahmadikia2024data,jadoon2025inverse,patel2025general}). Additionally, these frameworks can be employed with other NN-based frameworks for modeling the post-yield behavior \cite{meyer2023thermodynamically, jadoon2025automated, fuhg2023modular} to calibrate not only the yield but also the hardening behavior. While we developed our frameworks for materials exhibiting tension-compression symmetry in the yield function, these are equally applicable to the case of tension-compression asymmetry. Additionally, we introduced a method to ensure shape-consistent scaling of the yield functions, since our proposed frameworks are not degree-one homogeneous by design. However, this was done under the presumption of isotropic hardening. In case of distortional hardening, a future study can explore the possibility of not only modeling the initial yield, but also the evolution of the yield surface.

While the dataset used in this study may appear limited in size from a machine learning perspective, it is, in fact, one of the most comprehensive datasets available for characterizing plastic anisotropy from uniaxial tension experiments. In future work, we aim to extend our modeling efforts to capture the full uniaxial stress–strain response, beyond just the yield point, and to develop neural network–based elastoplastic models informed by full-field experimental data.

\section*{Acknowledgements}
This work was supported by the Laboratory Directed Research and Development program at Sandia National Laboratories, a multimission laboratory managed and operated by National Technology and Engineering Solutions of Sandia LLC, a wholly owned subsidiary of Honeywell International Inc. for the U.S. Department of Energy’s National Nuclear Security Administration under contract DE-NA0003525. This paper describes objective technical results and analysis. Any subjective views or opinions that might be expressed in the paper do not necessarily represent the views of the U.S. Department of Energy or the United States Government.

\clearpage
\bibliographystyle{unsrt}
\bibliography{references.bib}

\clearpage
\appendix
\renewcommand{\thetable}{\thesection.\arabic{table}}
\numberwithin{equation}{section}
\numberwithin{figure}{section}

\textbf{\large Appendices}

\section{Thermodynamic Considerations}\label{sec:Thermodynam}

\paragraph{Convexity of the Yield Function.}
Thermodynamic consistency of rate-independent plasticity requires the mechanical dissipation to be non-negative, i.e.,
\begin{equation}
    \mathcal{D}_{\text{mech}} = \bm{\sigma} : \dot{\bm{\varepsilon}}^p - k \dot{\kappa} \geq 0,
\end{equation}
where $\bm{\sigma}$ is the Cauchy stress, $\dot{\bm{\varepsilon}}^p$ is the plastic strain rate, $k$ is the thermodynamic force conjugate to the hardening variable $\kappa$, and $\dot{\kappa}$ is the corresponding rate. The yield function $f(\bm{\sigma}, k)$ defines the admissible elastic domain $\mathcal{E} := \{ (\bm{\sigma}, k)  \mid f(\bm{\sigma}, k) \leq 0 \}$. The support function of this convex set defines the dissipation potential:
\begin{equation}\label{eq:a2}
    \mathcal{D}(\dot{\bm{\varepsilon}}^p, \dot{\kappa}) = \sup_{(\bm{\sigma}, k) \in \mathcal{E}} \left\{ \bm{\sigma} : \dot{\bm{\varepsilon}}^p - k \dot{\kappa} \right\}.
\end{equation}
Assuming associative plasticity, the actual generalized forces $(\bm{\sigma}, -k)$ lie in the subdifferential of $\mathcal{D}$, i.e.,
\begin{equation}
    (\bm{\sigma}, -k) \in \partial \mathcal{D}(\dot{\bm{\varepsilon}}^p, \dot{\kappa}).
\end{equation}
By the definition of the subdifferential of a convex function, we have the inequality:
\begin{equation}\label{eq:a4}
    \mathcal{D}(\bm{\eta}, \zeta) \geq \mathcal{D}(\dot{\bm{\varepsilon}}^p, \dot{\kappa}) + \bm{\sigma} : (\bm{\eta} - \dot{\bm{\varepsilon}}^p) - k (\zeta - \dot{\kappa}) \quad \forall (\bm{\eta}, \zeta).
\end{equation}
Since the elastic domain $\mathcal{E}$ contains the origin (i.e., $(\bm{0}, 0) \in \mathcal{E}$), the supremum in Eq. \eqref{eq:a2} evaluated at $(\bm{0},0)$ yields zero:
\begin{equation}
    \mathcal{D}(\bm{0},0) = \sup_{(\bm{\sigma},k) \in \mathcal{E}} \{ \bm{\sigma} : 0 - k \cdot 0 \} = 0.
\end{equation}

Therefore, choosing $(\bm{\eta}, \zeta) = (\bm{0}, 0)$ and seeing that $\mathcal{D}(\bm{0}, 0) = 0$, we obtain:
\begin{equation}\label{eq:a5}
    \bm{\sigma} : \dot{\bm{\varepsilon}}^p - k \dot{\kappa} \geq \mathcal{D}(\dot{\bm{\varepsilon}}^p, \dot{\kappa}).
\end{equation}

From the definition of support functions, see \cref{rockafellar2015convex}, we also know that any support function over a convex set that contains the origin and is not a singleton must be nonnegative and strictly positive outside the origin. Therefore, we see that 
\begin{equation}
        \mathcal{D}(\dot{\bm{\varepsilon}}^p, \dot{\kappa}) \geq 0 .
\end{equation}

On the other hand, from the definition of $\mathcal{D}$ (and the definition of the supremum), it holds that
\begin{equation}\label{eq:a6}
    \bm{\sigma} : \dot{\bm{\varepsilon}}^p - k \dot{\kappa} \leq \mathcal{D}(\dot{\bm{\varepsilon}}^p, \dot{\kappa}) = \sup_{(\bm{\sigma}, k) \in \mathcal{E}} \left\{ \bm{\sigma} : \dot{\bm{\varepsilon}}^p - k \dot{\kappa} \right\}.
\end{equation}
Combining inequalities \eqref{eq:a5} and \eqref{eq:a6} yields equality, and we conclude that
\begin{equation}
     \bm{\sigma} : \dot{\bm{\varepsilon}}^p - k \dot{\kappa} = \mathcal{D}(\dot{\bm{\varepsilon}}^p, \dot{\kappa}) = \mathcal{D}_{\text{mech}} \geq 0,
    \label{A.7}
\end{equation}
which confirms that the dissipation inequality is satisfied. Importantly, this conclusion holds under the assumption that the dissipation potential $\mathcal{D}$ is convex. Since $\mathcal{D}$ is defined as the support function of the admissible elastic domain $\mathcal{E}$, its convexity is equivalent to the convexity of $\mathcal{E}$. Hence, the yield function $f(\bm{\sigma}, k)$ must be convex in both stress and hardening variables to ensure the convexity of $\mathcal{E}$ and, consequently, of $\mathcal{D}$. For more information, we refer to Han and Reddy \cite{han1999plasticity}.

It is crucial to emphasize that the convexity of $\mathcal{D}$ at a single point, e.g., around $(\bm{0}, 0)$, is not sufficient to ensure thermodynamic consistency. The dissipation inequality must hold for arbitrary loading directions $(\dot{\bm{\varepsilon}}^p, \dot{\kappa})$, including large finite increments encountered during plastic flow. Therefore, the subdifferential inequality \eqref{eq:a4} must hold for all $(\bm{\eta}, \zeta)$, which requires that $\mathcal{D}$ be globally convex. This global convexity ensures that the actual stress and internal variables can always be interpreted as subgradients of $\mathcal{D}$, maintaining consistency with associative flow rules and guaranteeing non-negative dissipation throughout arbitrary paths of plastic evolution.

\section{Deviatoric Plane}\label{section:dev_plane}

The deviatoric plane is commonly used to depict projections of the yield surface. Also referred to as $\pi\text{-plane}$, the deviatoric plane is a two-dimensional subspace within the three-dimensional principal stress space for pressure-independent yielding. The deviatoric plane lies orthogonal to the hydrostatic axis marked by the line where all principal stresses are equal. Here we outline the procedure for mapping to and from the deviatoric plane. First, given the Cauchy stress $\cauchy$, we obtain its deviator $\devcauchy := \text{dev}\left(\cauchy\right) = \cauchy - \frac{\text{tr}\left(\cauchy\right)}{3} \boldI$. We require an orthonormal basis $\{\pdir_1, \pdir_2, \pdir_3\}$ that will be used to obtain the associated scalar stress projections $\devpval_i = \pdir_i \cdot \devcauchy \pdir_i, \ i = 1, 2, 3$. When this basis corresponds to the principal directions (the eigenbasis), the projections are the principal deviatoric stresses (the eigenvalues). They are related to the principal stresses through a shift factor $\pval_i = \devpval_i + \frac{\text{tr}\left(\cauchy\right)}{3}$. 

We note that information will be lost (i.e., we cannot completely recover $\devcauchy$ from the scalar projections $\devpval_i$ and basis vectors $\{\pdir_1, \pdir_2, \pdir_3\}$) when the orthonormal basis used for projection does not coincide with the principal directions. At a given material point in a deforming body, the principal directions will generally change over the course of the deformation. One noteworthy exception is a uniaxial tension experiment. Nevertheless, we can plot the projections $\devpval_i$ in a 3D right-handed coordinate system using the $\{\pdir_1, \pdir_2, \pdir_3\}$ basis. The transformation to the deviatoric plane requires two rotations, i.e., a rotation of $\theta_1 = \frac{\pi}{4}$ about the $\devpval_2$ axis to obtain $(\devpval_1', \devpval_2', \devpval_3')$, followed by a rotation of $\theta_2 = -\cos^{-1} \sqrt{\frac{2}{3}}$ about the $\devpval_1'$ axis which yields $(\devpval_1'', \devpval_2'', \devpval_3'')$ \cite{borja_plasticity_2013}. By construction, $\devpval_3'' = 0$, thus the deviatoric plane coordinates $(\pi_1, \pi_2)$ are obtained by dropping $\devpval_3''$ and scaling by $\sqrt{\frac{3}{2}}$:

\begin{align}
\left\{ \begin{array}{c}
\pi_1 \\
\pi_2
\end{array}
\right\} 
&= 
\begin{bmatrix}
\sqrt{\frac{3}{2}} & 0 & 0 \\
0 & \sqrt{\frac{3}{2}} & 0
\end{bmatrix}
\begin{bmatrix}
1 & 0 & 0 \\
0 & \cos \theta_2 & \sin \theta_2 \\
0 & -\sin \theta_2 & \cos \theta_2
\end{bmatrix}
\begin{bmatrix}
\cos \theta_1 & 0 & -\sin \theta_1 \\
0 & 1 & 0 \\
\sin \theta_1 & 0 & \cos \theta_1
\end{bmatrix}
\left\{ \begin{array}{c}
\devpval_1 \\
\devpval_2 \\
\devpval_3
\end{array}
\right\} \\
&=
\begin{bmatrix}
\sqrt{\frac{3}{2}} & 0 & 0 \\
0 & \sqrt{\frac{3}{2}} & 0
\end{bmatrix}
\begin{bmatrix}
1 & 0 & 0 \\
0 & \sqrt{\frac{2}{3}} & -\sqrt{\frac{1}{3}} \\
0 & \sqrt{\frac{1}{3}} & \sqrt{\frac{2}{3}}
\end{bmatrix}
\begin{bmatrix}
\sqrt{\frac{1}{2}} & 0 & -\sqrt{\frac{1}{2}} \\
0 & 1 & 0 \\
\sqrt{\frac{1}{2}} & 0 & \sqrt{\frac{1}{2}}
\end{bmatrix}
\left\{ \begin{array}{c}
\devpval_1 \\
\devpval_2 \\
\devpval_3
\end{array}
\right\} \\
&=
\begin{bmatrix}
\sqrt{\frac{3}{2}} & 0 & 0 \\
0 & \sqrt{\frac{3}{2}} & 0
\end{bmatrix}
\begin{bmatrix}
\sqrt{\frac{1}{2}} & 0 & -\sqrt{\frac{1}{2}} \\
-\sqrt{\frac{1}{6}} & \sqrt{\frac{2}{3}} & -\sqrt{\frac{1}{6}} \\
\sqrt{\frac{1}{3}} & \sqrt{\frac{1}{3}} & \sqrt{\frac{1}{3}}
\end{bmatrix}
\left\{ \begin{array}{c}
\devpval_1 \\
\devpval_2 \\
\devpval_3
\end{array}
\right\} \\
&=
\begin{bmatrix}
\sqrt{\frac{2}{3}} & 0 & -\sqrt{\frac{2}{3}}  \\
-\frac{1}{2} & 1 & -\frac{1}{2}
\end{bmatrix}
\left\{ \begin{array}{c}
\devpval_1 \\
\devpval_2 \\
\devpval_3
\end{array}
\right\}.
\end{align}

This scaling yields a correspondence between the unit circle in the deviatoric plane and the parameterization of $(\devpval_1, \devpval_2, \devpval_3)$ defined by

\begin{align}
(\devpval_1, \devpval_2, \devpval_3) &= \frac{2}{3} \left(\cos \psi, \cos \left(\psi - \frac{2 \pi}{3}\right),
\cos \left(\psi + \frac{2 \pi}{3}\right)\right)
\label{eq:psi_angles} \quad \text{with} \quad \psi = [0, 2 \pi)\, .
\end{align}

The polar angle in the deviatoric plane $\alpha = \arctan\left({\pi_2 / \pi_1}\right)$ is offset such that $\alpha = \psi - \frac{\pi}{6}$. The combination of scaling and rotation provides a convenient link between uniaxial stress states (under certain projections) and the deviatoric projected stress scalars. For example, if we assume the projection basis is the standard basis, the uniaxial stress state

\begin{equation}
[\cauchy] =
\begin{bmatrix}
1 & 0 & 0 \\
0 & 0 & 0 \\
0 & 0 & 0
\end{bmatrix}
\end{equation}

results in the deviator

\begin{equation}
[\devcauchy] =
\begin{bmatrix}
\frac{2}{3} & 0 & 0 \\
0 & -\frac{1}{3} & 0 \\
0 & 0 & -\frac{1}{3}
\end{bmatrix},
\end{equation}

which corresponds to $\psi = 0$ in \eqref{eq:psi_angles}. This is the point on the unit circle (i.e.\ $r = 1$) with polar angle $\alpha = -\frac{\pi}{6}$. Furthermore, the unit circle in the deviatoric plane will correspond to deviatoric stress tensors with unit $J_2$ effective stress when the principal directions and projection basis coincide. Similarly, we can also construct the mapping from the deviatoric plane to the projected deviatoric stress scalars as:

\begin{align}
\left\{ \begin{array}{c}
\devpval_1 \\
\devpval_2 \\
\devpval_3
\end{array}
\right\}
&=
\begin{bmatrix}
\cos \theta_1 & 0 & -\sin \theta_1 \\
0 & 1 & 0 \\
\sin \theta_1 & 0 & \cos \theta_1
\end{bmatrix}^T
\begin{bmatrix}
1 & 0 & 0 \\
0 & \cos \theta_2 & \sin \theta_2 \\
0 & -\sin \theta_2 & \cos \theta_2
\end{bmatrix}^T
\begin{bmatrix}
\sqrt{\frac{2}{3}} & 0 \\
0 & \sqrt{\frac{2}{3}} \\
0 & 0
\end{bmatrix}
\left\{ \begin{array}{c}
\pi_1 \\
\pi_2
\end{array}
\right\} \\
&=
\begin{bmatrix}
\sqrt{\frac{1}{2}} & 0 & \sqrt{\frac{1}{2}} \\
0 & 1 & 0 \\
-\sqrt{\frac{1}{2}} & 0 & \sqrt{\frac{1}{2}}
\end{bmatrix}
\begin{bmatrix}
1 & 0 & 0 \\
0 & \sqrt{\frac{2}{3}} & \sqrt{\frac{1}{3}} \\
0 & -\sqrt{\frac{1}{3}} & \sqrt{\frac{2}{3}}
\end{bmatrix}
\begin{bmatrix}
\sqrt{\frac{2}{3}} & 0 \\
0 & \sqrt{\frac{2}{3}} \\
0 & 0
\end{bmatrix}
\left\{ \begin{array}{c}
\pi_1 \\
\pi_2
\end{array}
\right\} \\
&=
\begin{bmatrix}
\sqrt{\frac{1}{2}} & -\sqrt{\frac{1}{6}} & \sqrt{\frac{1}{3}} \\
0 & \sqrt{\frac{2}{3}} & \sqrt{\frac{1}{3}} \\
 -\sqrt{\frac{1}{2}} & -\sqrt{\frac{1}{6}} & \sqrt{\frac{1}{3}}
\end{bmatrix}
\begin{bmatrix}
\sqrt{\frac{2}{3}} & 0 \\
0 & \sqrt{\frac{2}{3}} \\
0 & 0
\end{bmatrix}
\left\{ \begin{array}{c}
\pi_1 \\
\pi_2
\end{array}
\right\} \\
&=
\begin{bmatrix}
\sqrt{\frac{1}{3}} & -\frac{1}{3}\\
0 & \frac{2}{3}  \\
-\sqrt{\frac{1}{3}} & -\frac{1}{3}
\end{bmatrix}
\left\{ \begin{array}{c}
\pi_1 \\
\pi_2
\end{array}
\right\}.
\end{align}

\setcounter{table}{0}
\section{Parameter study} \label{app:paramStudy}
In this section, we investigate the influence of the number of trainable parameters on our frameworks. Since the number of trainable parameters does not change for the phenomenological models, they are not considered in this study, and only the NN-based approaches are examined. Apart from the trainable parameter set presented in Table \ref{tab:no_parameters}, we create two additional sets of trainable parameters as reported in Table \ref{tab:no_parameters_ParamStudy}. These sets are denoted as lower and higher, and have a lower and higher number of trainable parameters than the original set of Table \ref{tab:no_parameters}. We remark that we choose a comparable number of trainable parameters for all frameworks. We use $\alpha=10$ in Eq. \ref{eq:earlyStopCriteria}. 

\begin{table}[h!]
\centering
\caption{Number of trainable parameters for each approach.}
\label{tab:no_parameters_ParamStudy}
\begin{tabular}{lcc}
\toprule
\multirow{2}{*}{Method} 
  & \multicolumn{2}{c}{Trainable parameters} \\
\cmidrule(lr){2-3}
  & Lower & Higher \\ 
\midrule
ICNN                  & 135 & 601 \\
Hybrid                & 141 & 607 \\
$\text{PI-ICNN}_{1}$  & 132 & 632 \\
$\text{PI-ICNN}_{2}$  & 142 & 656 \\
\bottomrule
\end{tabular}
\end{table}

\begin{figure}[htbp]
  \centering
  \begin{subfigure}{0.48\textwidth}
    \centering
    \includegraphics[width=\linewidth]{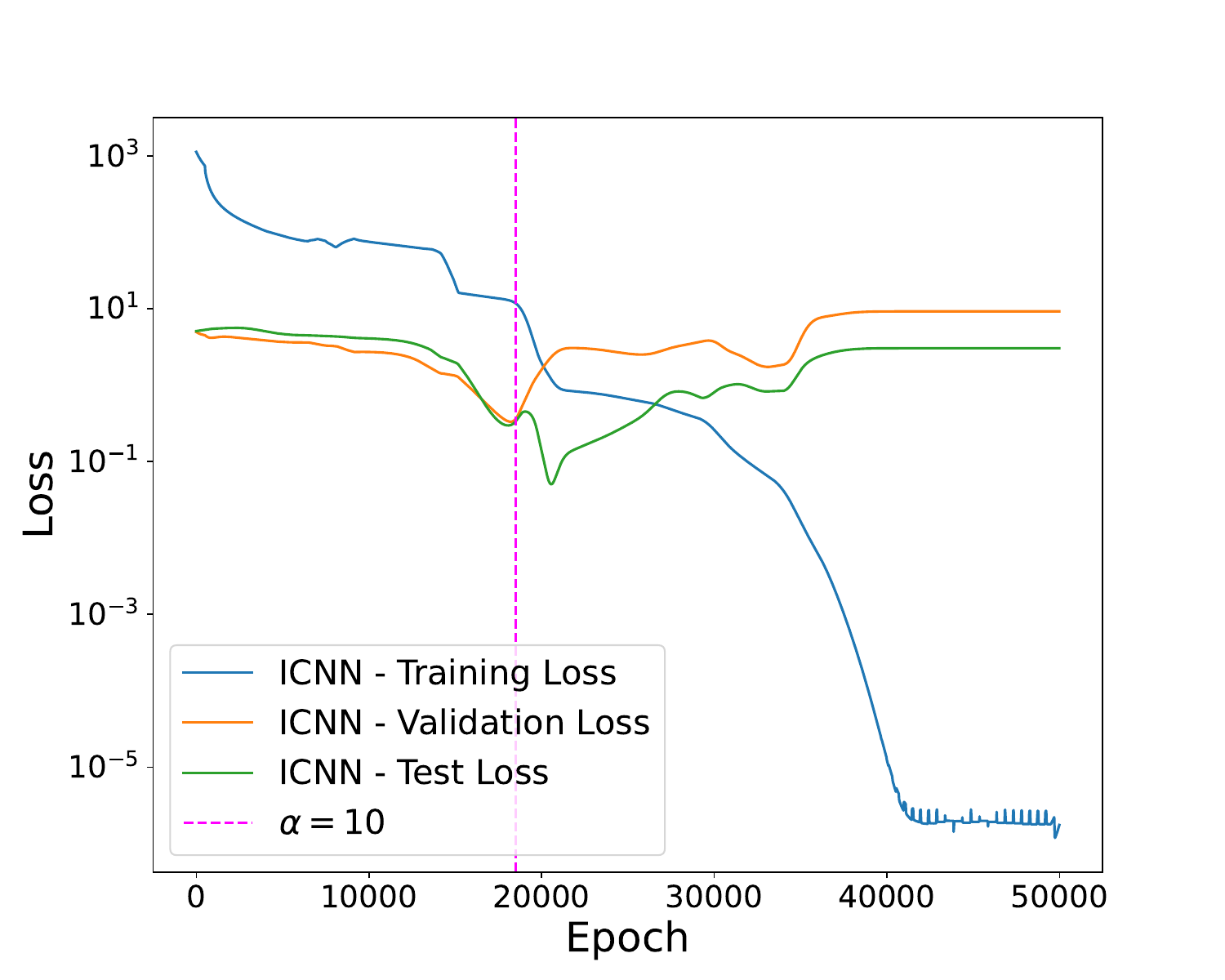}
    \caption{}
    \label{fig:loss_icnn_lowParams}
  \end{subfigure}\hfill
  \begin{subfigure}{0.48\textwidth}
    \centering
    \includegraphics[width=\linewidth]{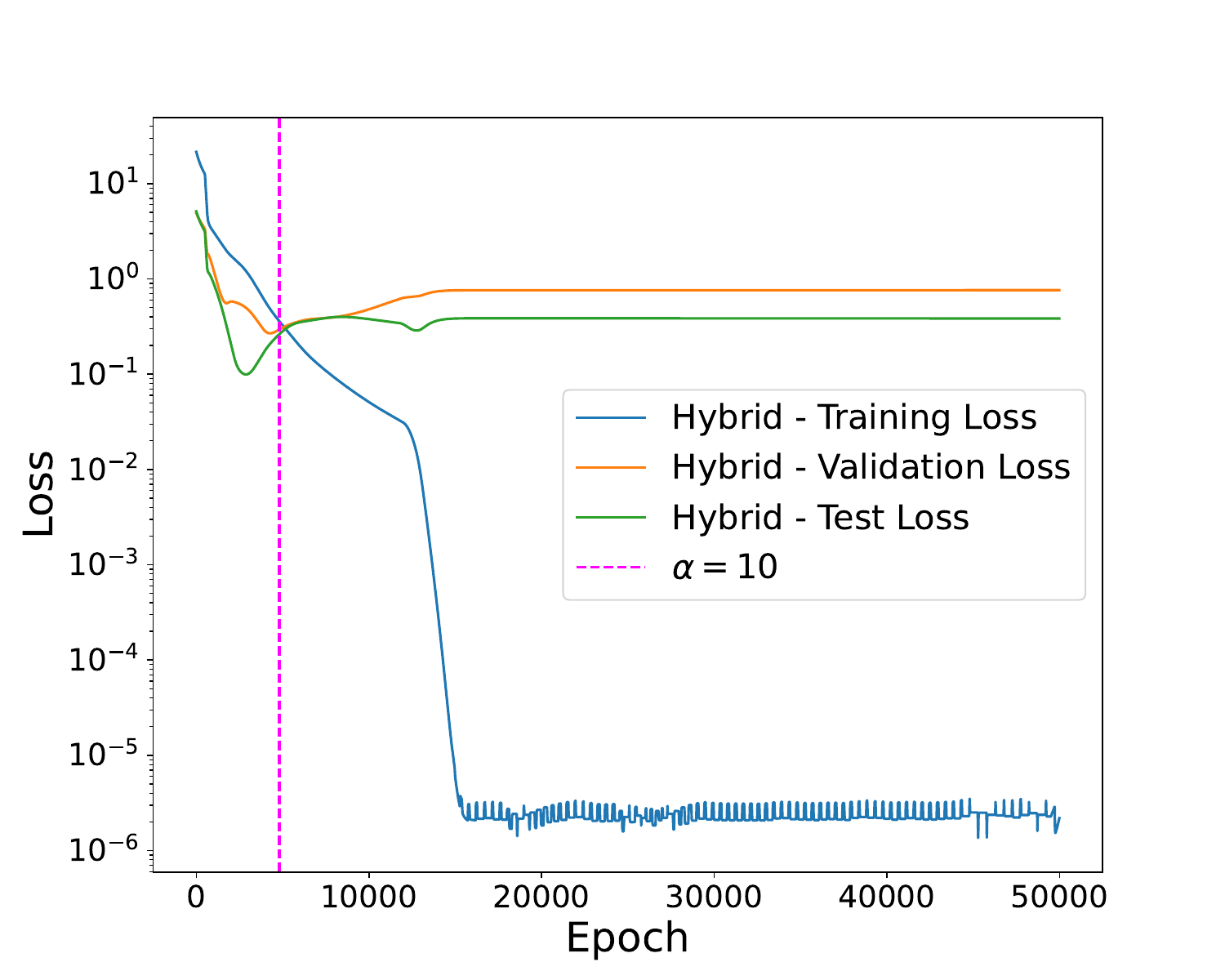}
    \caption{}
    \label{fig:loss_hybrid_lowParams}
  \end{subfigure}

  \caption{Loss evolution for (a) ICNN and (b) hybrid models with lower number of trainable parameters as reported in Table \ref{tab:no_parameters_ParamStudy}.}
  \label{fig:loss_icnn_hybrid_lowerParams}
\end{figure}

The error metrics for the training, validation, and test data are presented in Tables \ref{tab:Lower_train}, \ref{tab:Lower_validation}, and \ref{tab:Lower_test}. We see a similar trend in the validation and test error metrics as compared to the results reported in Section \ref{Sec:Results}. However, ICNNs now also perform poorly on the training data. Upon an inspection of the loss evolution over training epochs, it was observed that ICNN on its own had a high loss at initialization and consequently started to overfit at a higher value of training loss as compared to the hybrid approach, which imparted some information from Hill-48 into the yield function, resulting in lower losses at the initialization. This led to delayed overfitting, or at a relatively lower training loss, as compared to pure ICNNs. This is shown in Figure \ref{fig:loss_icnn_hybrid_lowerParams} for dataset 3, and a similar trend in loss evolution was observed for the other datasets as well. 

\begin{table}[h!]
\centering
\caption{Error metrics for the training dataset with lower number of trainable parameters}
\label{tab:Lower_train}
\begin{tabular}{lcccc}
\toprule
Method & Max $|f|$ & Mean $|f|$ & Max $|\Delta r|$ & Mean $|\Delta r|$ \\
\midrule
ICNN         & {257.91} & {100.16} & 0.78 & 0.20 \\
Hybrid       & \textbf{24.05}  & \textbf{8.22}  & 0.34 & 0.11 \\
$\text{PI-ICNN}_1$    & 37.68  & 17.14  & 0.17 & \textbf{0.05} \\
$\text{PI-ICNN}_2$    & 33.48 & 12.06 & \textbf{0.13} & \textbf{0.05} \\
\bottomrule
\end{tabular}
\end{table}

\begin{table}[h!]
\centering
\caption{Error metrics for the validation dataset with lower number of trainable parameters}
\label{tab:Lower_validation}
\begin{tabular}{lcccc}
\toprule
Method & Max $|f|$ & Mean $|f|$ & Max $|\Delta r|$ & Mean $|\Delta r|$ \\
\midrule
ICNN         & 141.75  & 108.98  & 0.64 & 0.39 \\
Hybrid       & 162.84  & 113.59  & 0.73 & 0.46 \\
$\text{PI-ICNN}_1$    & \textbf{23.76}  & \textbf{17.90}  & 0.30 & 0.19 \\
$\text{PI-ICNN}_2$    & 44.81  & 34.68 & \textbf{0.24} & \textbf{0.17} \\
\bottomrule
\end{tabular}
\end{table}

\begin{table}[h!]
\centering
\caption{Error metrics for the test dataset with lower number of trainable parameters}
\label{tab:Lower_test}
\begin{tabular}{lcccc}
\toprule
Method & Max $|f|$ & Mean $|f|$ & Max $|\Delta r|$ & Mean $|\Delta r|$ \\
\midrule
ICNN         & 216.44  & 145.46  & 1.04 & 0.62 \\
Hybrid       & 176.16  & 121.31  & 0.98 & 0.59 \\
$\text{PI-ICNN}_1$    & 36.47  & 26.29  & \textbf{0.30} & \textbf{0.18} \\
$\text{PI-ICNN}_2$    & \textbf{33.32} & \textbf{20.72} & 0.37 & 0.22 \\
\bottomrule
\end{tabular}
\end{table}

Increasing the number of trainable parameters did not seem to influence the behavior of any of the frameworks, and similar trends were observed for the training, test, and validation losses as compared to the original set reported in Section \ref{Sec:Results}. The respective metrics are reported in Tables \ref{tab:Higher_train}-\ref{tab:Higher_test}.

\begin{table}[h!]
\centering
\caption{Error metrics for the training dataset with higher number of trainable parameters}
\label{tab:Higher_train}
\begin{tabular}{lcccc}
\toprule
Method & Max $|f|$ & Mean $|f|$ & Max $|\Delta r|$ & Mean $|\Delta r|$ \\
\midrule
ICNN         & 38.50 & 14.37 & 0.20 & \textbf{0.05} \\
Hybrid       & \textbf{25.31}  & \textbf{11.54}  & 0.33 & 0.11 \\
$\text{PI-ICNN}_1$    & 40.75  & 16.81  & \textbf{0.19} & \textbf{0.05} \\
$\text{PI-ICNN}_2$    & 39.81 & 14.79 & 0.26 & 0.06 \\
\bottomrule
\end{tabular}
\end{table}

\begin{table}[h!]
\centering
\caption{Error metrics for the validation dataset with higher number of trainable parameters}
\label{tab:Higher_validation}
\begin{tabular}{lcccc}
\toprule
Method & Max $|f|$ & Mean $|f|$ & Max $|\Delta r|$ & Mean $|\Delta r|$ \\
\midrule
ICNN         & 137.80  & 98.75  & 0.60 & 0.36 \\
Hybrid       & 139.46  & 111.69  & 0.60 & 0.36 \\
$\text{PI-ICNN}_1$    & \textbf{38.93}  & \textbf{29.48}  & \textbf{0.31} & \textbf{0.21} \\
$\text{PI-ICNN}_2$    & 46.40  & 34.42 & 0.36 & 0.23 \\
\bottomrule
\end{tabular}
\end{table}

\begin{table}[h!]
\centering
\caption{Error metrics for the test dataset with higher number of trainable parameters}
\label{tab:Higher_test}
\begin{tabular}{lcccc}
\toprule
Method & Max $|f|$ & Mean $|f|$ & Max $|\Delta r|$ & Mean $|\Delta r|$ \\
\midrule
ICNN         & 135.12  & 100.18  & 0.87 & 0.50 \\
Hybrid       & 163.70  & 102.89  & 0.82 & 0.52 \\
$\text{PI-ICNN}_1$    & 42.07  & 30.85  & \textbf{0.31} & \textbf{0.20} \\
$\text{PI-ICNN}_2$    & \textbf{38.86} & \textbf{27.58} & 0.45 & 0.25 \\
\bottomrule
\end{tabular}
\end{table}

\setcounter{table}{0}
\section{Results for the lowest validation loss epoch} \label{app:alphaequal0}

The choice of $\alpha\ (\alpha=10)$ in Eq. \eqref{eq:earlyStopCriteria} was empirical. Therefore, we also report results for $\alpha=0$, which translates to the epoch with the lowest validation loss. Since phenomenological models show no signs of overfitting, their metrics are replicated from Section \ref{Sec:Results} and presented here to make a comparison. The number of trainable parameters for the results reported here corresponds to those presented in Table \ref{tab:no_parameters}. The error metrics are presented in Tables \ref{tab:alpha0_train}-\ref{tab:alpha0_test}. They follow a similar trend as observed in the results reported in Section \ref{Sec:Results}.

\begin{table}[h!]
\centering
\caption{Error metrics for the training dataset with $\alpha=0$ in Eq. \eqref{eq:earlyStopCriteria}}
\label{tab:alpha0_train}
\begin{tabular}{lcccc}
\toprule
Method & Max $|f|$ & Mean $|f|$ & Max $|\Delta r|$ & Mean $|\Delta r|$ \\
\midrule
Hill-48         & 90.14 & 42.25 & 0.99 & 0.29 \\
Yld2004-18p       & \textbf{13.23} & \textbf{5.55}  & 0.25 & \textbf{0.04} \\
ICNN         & 61.77 & 25.85 & 0.29 & 0.09 \\
Hybrid       & 51.92  & 17.88  & 0.42 & 0.13 \\
$\text{PI-ICNN}_1$    & 40.46  & 17.90  & 0.18 & 0.06 \\
$\text{PI-ICNN}_2$    & 21.97  & 8.35 & \textbf{0.11} & \textbf{0.04} \\
\bottomrule
\end{tabular}
\end{table}

\begin{table}[h!]
\centering
\caption{Error metrics for the validation dataset with $\alpha=0$ in Eq. \eqref{eq:earlyStopCriteria}}
\label{tab:alpha0_validation}
\begin{tabular}{lcccc}
\toprule
Method & Max $|f|$ & Mean $|f|$ & Max $|\Delta r|$ & Mean $|\Delta r|$ \\
\midrule
Hill-48         & 71.95 & 51.87 & 0.46 & 0.31 \\
Yld2004-18p       & 48.58  & 31.19  & 2.51 & 1.34 \\
ICNN         & 131.56  & 87.57  & 0.48 & 0.31 \\
Hybrid       & 181.02  & 123.04  & 0.58 & 0.42 \\
$\text{PI-ICNN}_1$    & \textbf{23.82}  & \textbf{17.96}  & \textbf{0.30} & \textbf{0.19} \\
$\text{PI-ICNN}_2$    & 33.47  & 23.32  & 0.36 & 0.21 \\
\bottomrule
\end{tabular}
\end{table}

\begin{table}[h!]
\centering
\caption{Error metrics for the test dataset with $\alpha=0$ in Eq. \eqref{eq:earlyStopCriteria}}
\label{tab:alpha0_test}
\begin{tabular}{lcccc}
\toprule
Method & Max $|f|$ & Mean $|f|$ & Max $|\Delta r|$ & Mean $|\Delta r|$ \\
\midrule
Hill-48         & 57.09 & 45.88 & 1.50 & 0.91 \\
Yld2004-18p       & 48.03 & 33.32 & 1.04 & 0.59 \\
ICNN         & 172.66  & 118.31  & 1.00 & 0.61 \\
Hybrid       & 213.75  & 140.97  & 0.94 & 0.63 \\
$\text{PI-ICNN}_1$    & 39.59  & 28.72  & \textbf{0.31} & \textbf{0.19} \\
$\text{PI-ICNN}_2$    & \textbf{38.84}  & \textbf{24.32} & 0.56 & 0.31 \\
\bottomrule
\end{tabular}
\end{table}
\setcounter{table}{0}
\section{Training on the entire dataset} \label{app:completeDataset}
Experimental anisotropic yield data is generally scarce, and training overparameterized regression models on a small dataset always runs the risk of overfitting. Therefore, it is necessary to split the dataset into training, validation, and testing data to ensure our models generalize well. However, this split is not commonly applied when training phenomenological models (even though they might be overparameterized as well). Here we report each framework's performance when trained on our entire dataset, without splitting it into validation and test sets. To evaluate their robustness, we generate three different parameter initializations for each framework, with the number of trainable parameters corresponding to those reported in Table \ref{tab:no_parameters}. Results for Hill-48 and Yld2004-18p are the same as those reported in Table \ref{tab:cmaes_corona} since they are also averaged over three different initializations. The average error metrics for these phenomenological models, as well as the NN-based approaches, are reported in Table \ref{tab:entireDataset}.

\begin{table}[h!]
\centering
\caption{Average error metrics for training over the entire dataset.}
\label{tab:entireDataset}
\begin{tabular}{lcccc}
\toprule
Method & Max $|f|$ & Mean $|f|$ & Max $|\Delta r|$ & Mean $|\Delta r|$ \\
\midrule
Hill-48 & 86.57 & 39.53 & 0.90 & 0.22 \\
Yld2004-18p & 14.93 & 6.36  & 0.16 & 0.04 \\
ICNN         & \textbf{4e-3} & \textbf{2e-3} & \textbf{6e-5} & \textbf{2e-5} \\
Hybrid       & \textbf{6e-3}  & \textbf{4e-3}  & \textbf{8e-5} & \textbf{1e-5} \\
$\text{PI-ICNN}_1$    & 28.84  & 12.02  & 0.07 & 0.03 \\
$\text{PI-ICNN}_2$    & 10.41 & 4.52 & 0.01 & 3e-3 \\
\bottomrule
\end{tabular}
\end{table}
We can see that, even when trained on the complete dataset, Hill-48 is unable to accurately predict the yield and Lankford coefficients. In comparison, Yld2004-18p yields significantly lower errors. Both $\text{PI-ICNN-based}$ approaches show comparable results. Both show a better r-value fit than Yld2004-18p, and $\text{PI-ICNN}_2$ also predicts yield better. The ICNN and the hybrid approach fit the data (almost) perfectly,  albeit (most likely) at the expense of overfitting and loss of generalization. 
\setcounter{table}{0}
\section{Loss evolution} \label{app:loss_evolution}
Since it is not possible to report the loss evolution for all the datasets of Table \ref{tab:dataset_splits}, we only present the loss evolution for the randomly chosen dataset 3. We believe this to be representative.

\begin{figure}[htbp]
  \centering
  \begin{subfigure}{0.48\textwidth}
    \centering
    \includegraphics[width=\linewidth]{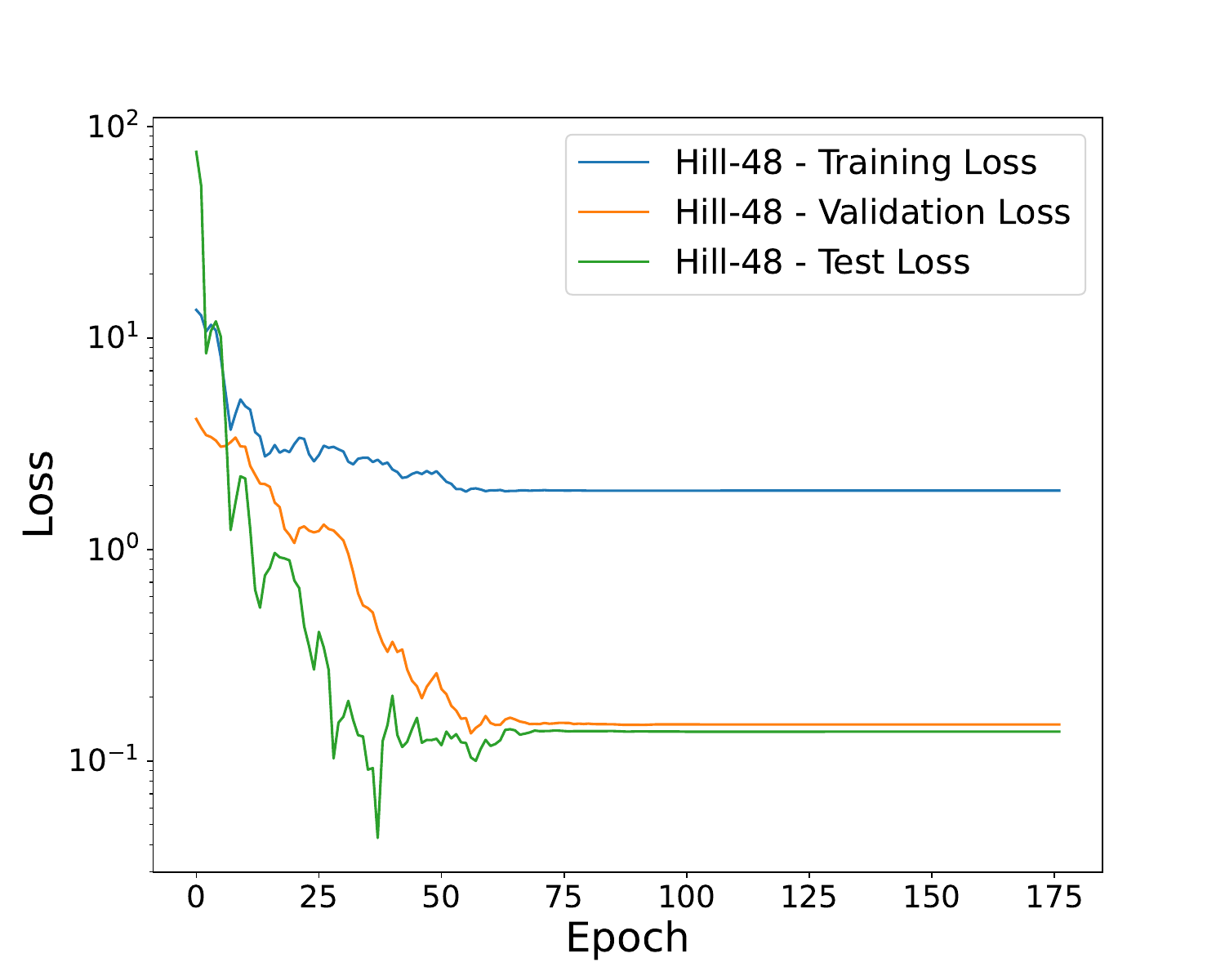}
    \caption{}
    \label{fig:loss_hill}
  \end{subfigure}\hfill
  \begin{subfigure}{0.48\textwidth}
    \centering
    \includegraphics[width=\linewidth]{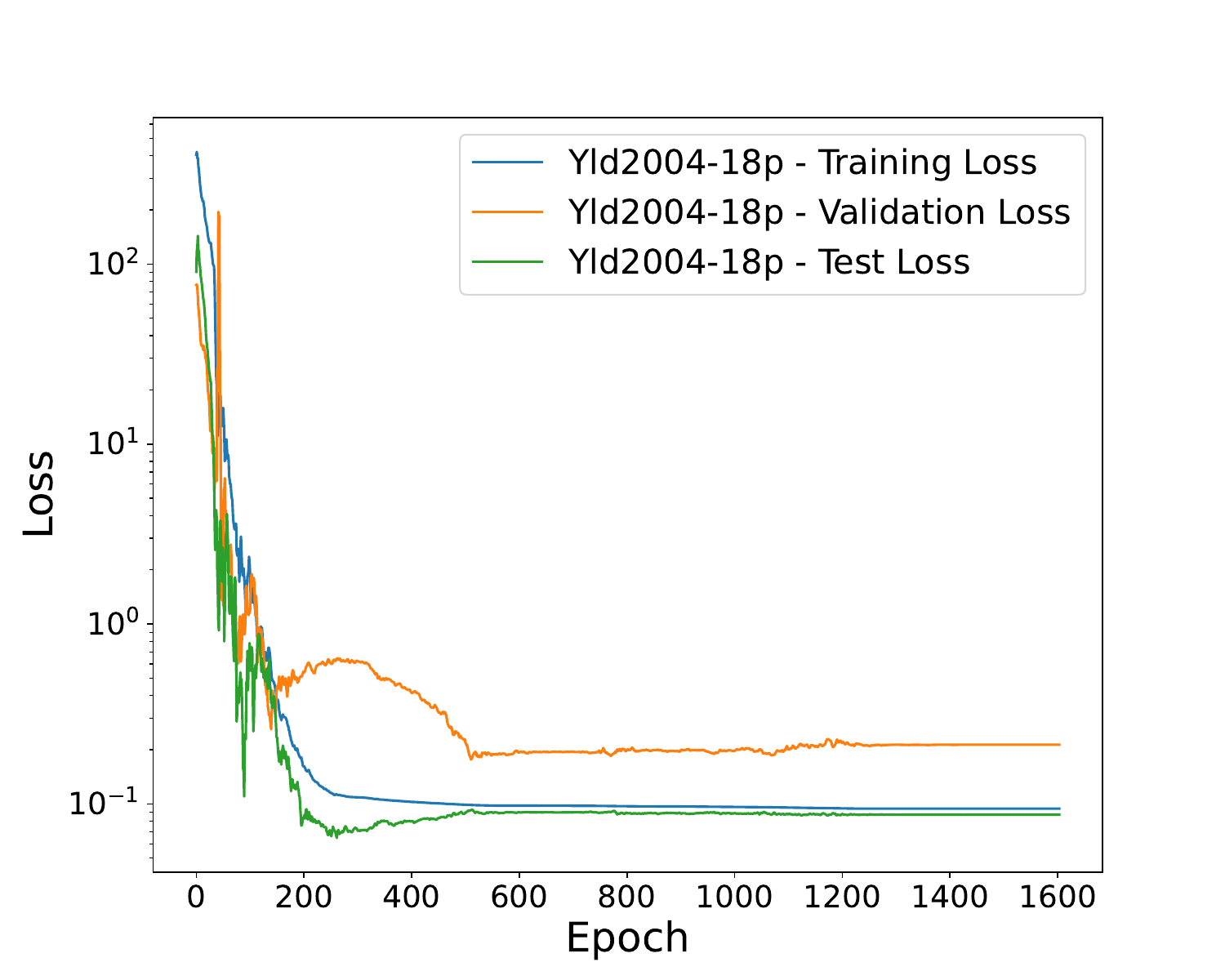}
    \caption{}
    \label{fig:loss_barlat}
  \end{subfigure}

  \vspace{1ex}  

  \begin{subfigure}{0.48\textwidth}
    \centering
    \includegraphics[width=\linewidth]{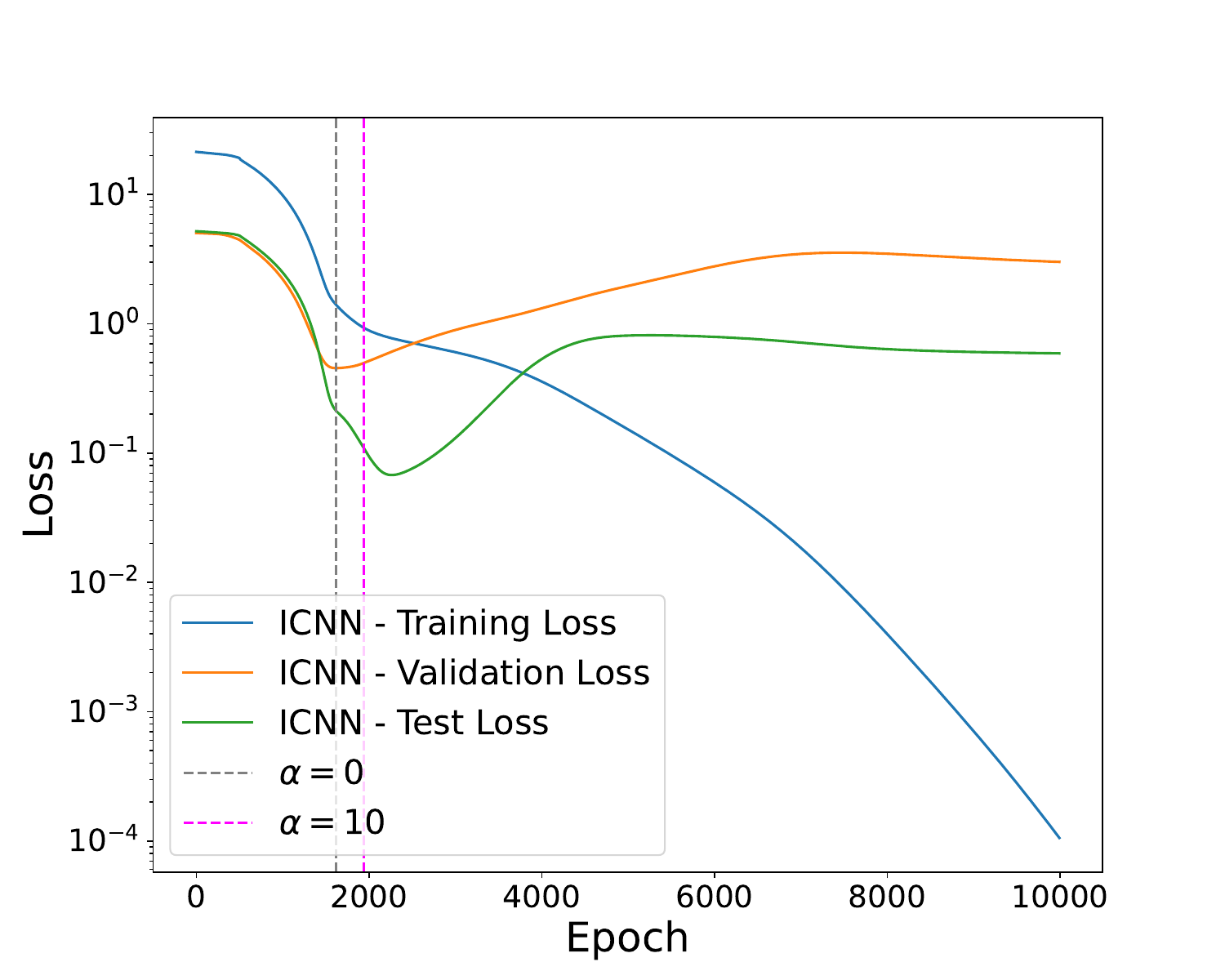}
    \caption{}
    \label{fig:loss_icnn}
  \end{subfigure}\hfill
  \begin{subfigure}{0.48\textwidth}
    \centering
    \includegraphics[width=\linewidth]{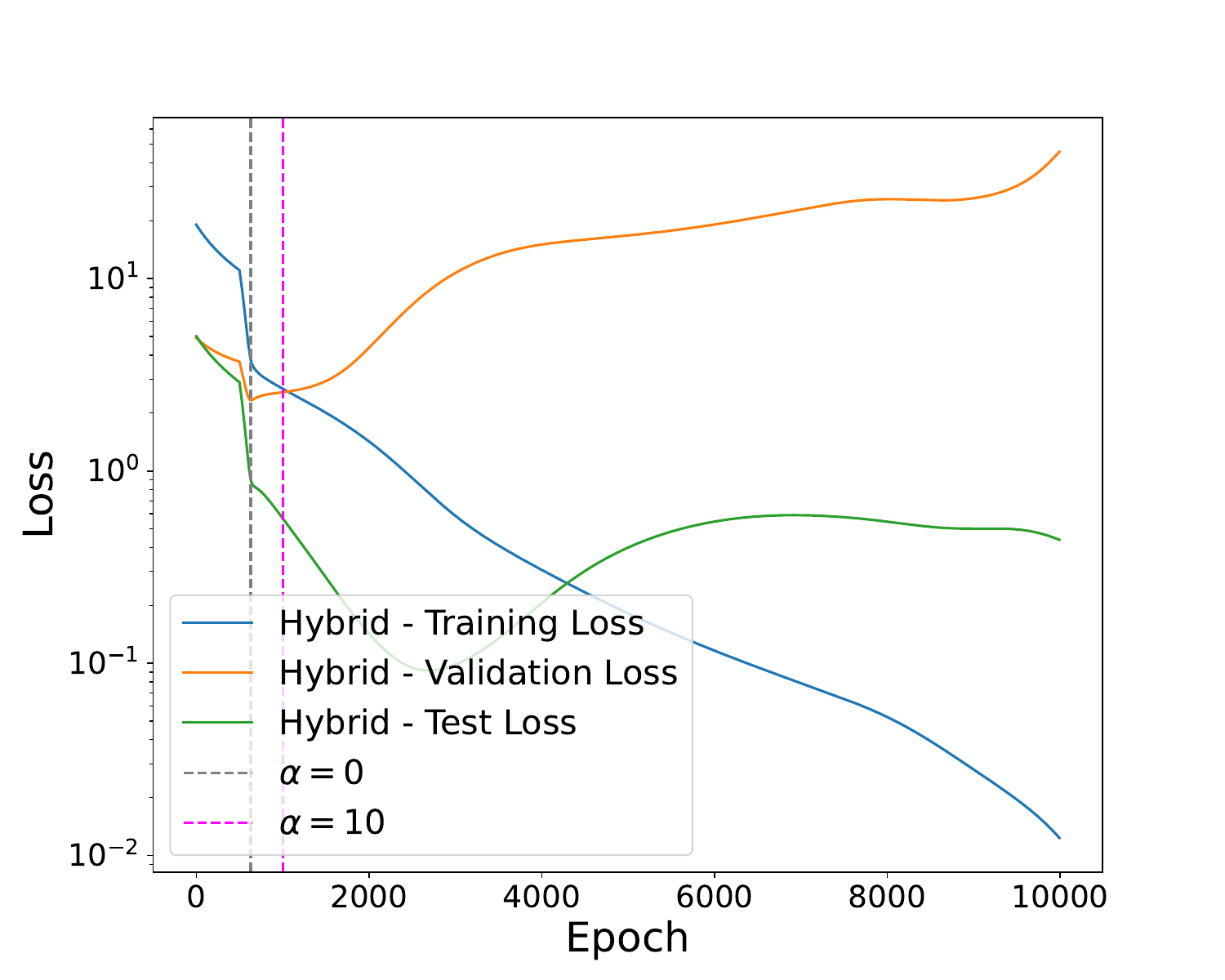}
    \caption{}
    \label{fig:loss_hybrid}
  \end{subfigure}

  \vspace{1ex}

  \begin{subfigure}{0.48\textwidth}
    \centering
    \includegraphics[width=\linewidth]{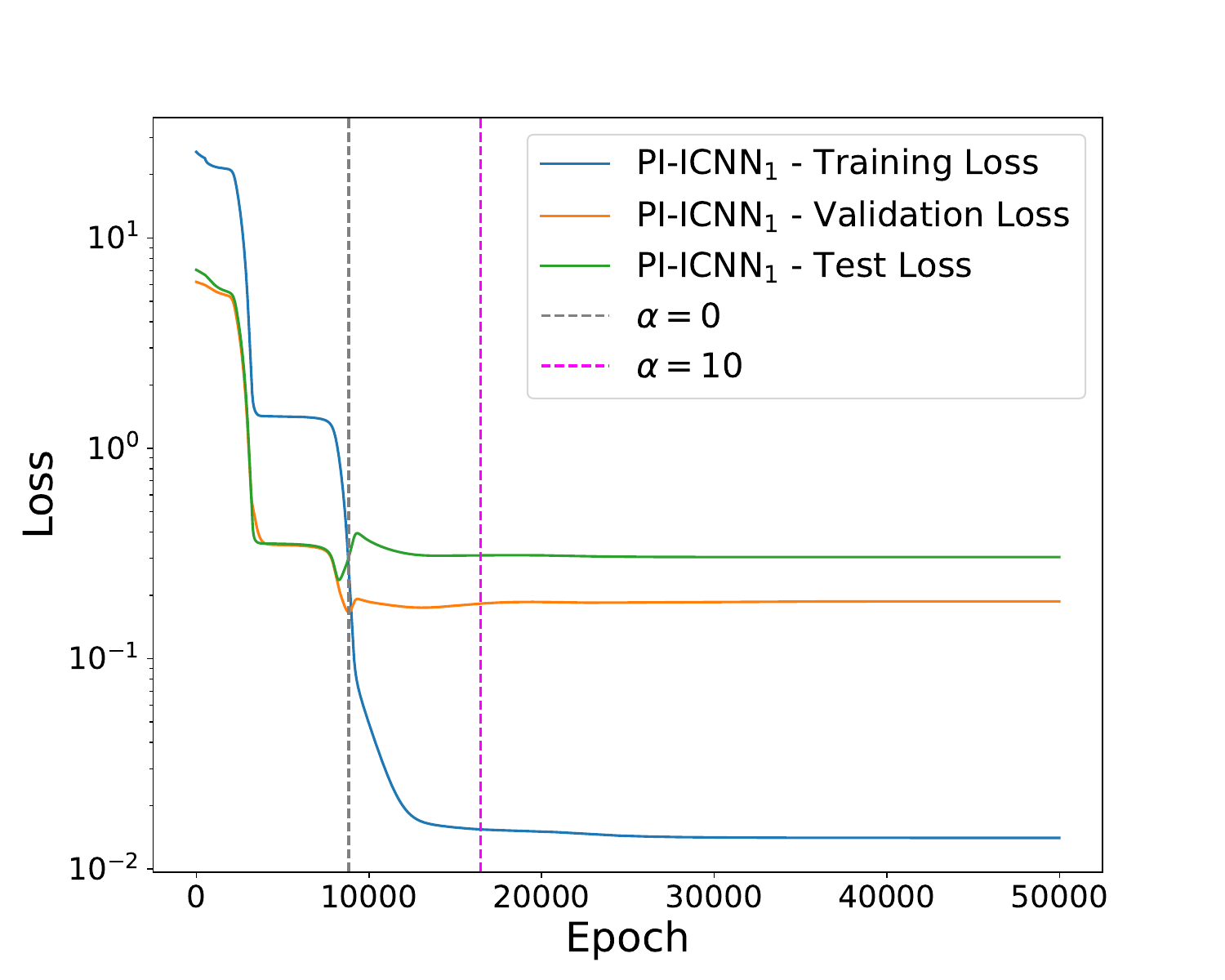}
    \caption{}
    \label{fig:loss_picnn1}
  \end{subfigure}\hfill
  \begin{subfigure}{0.48\textwidth}
    \centering
    \includegraphics[width=\linewidth]{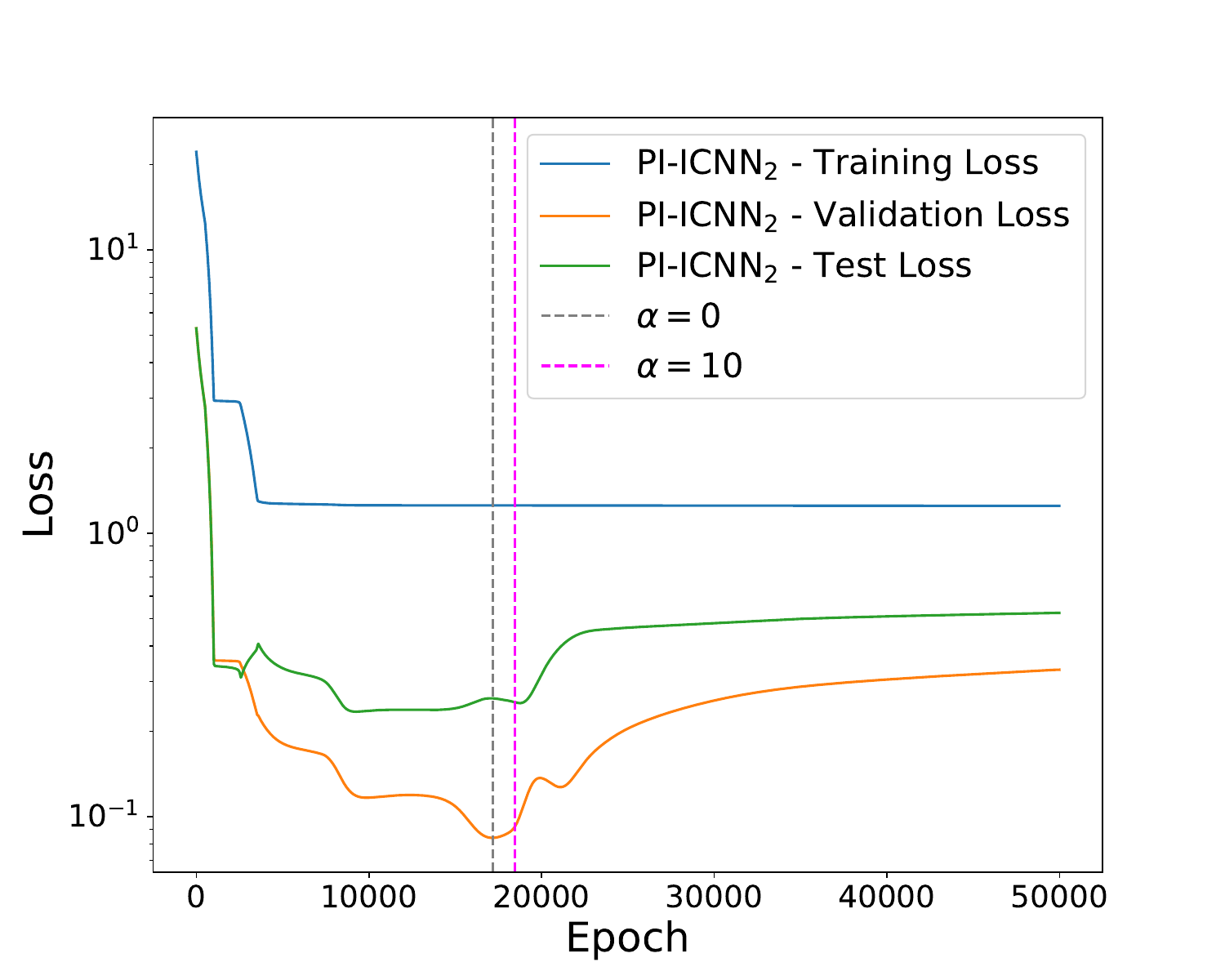}
    \caption{}
    \label{fig:loss_picnn2}
  \end{subfigure}

  \caption{Loss evolution for each framework used in this study for dataset 3. The plots show separately the evolution of training, validation and test losses along with vertical dashed lines marking the stoppage epochs in case an early stopping strategy is employed.}
  \label{fig:loss_all}
\end{figure}

The losses are reported in Figure \ref{fig:loss_all}.  Figures \ref{fig:loss_hill} and \ref{fig:loss_barlat} show that the phenomenological models do not experience overfitting. Therefore, an early stopping criterion would yield the same results. However, since we use an evolutionary algorithm for these models, their loss evolution exhibits a lot of noise, especially at the earlier epochs. To avoid complications, we therefore deliberately do not report early stopping criterion values. 

For the NN-based frameworks, we report losses that are smoothened by a simple moving average over a window of 2000 epochs. Although all the networks are trained for 50,000 epochs, we only show the loss evolution up to epoch 10,000 for the ICNN and hybrid framework in Figures \ref{fig:loss_icnn} and \ref{fig:loss_hybrid}. This is done to better visualize the overfitting phenomenon since the training losses keep decreasing over the next epochs, making it difficult to appreciate the overfitting (even in y-log-scale). Nevertheless, we see that the validation loss already starts increasing in the initial stages of training for both these approaches, and the stopping criterion activates quite early. The dashed lines, as can be seen in the legend, mark the epochs chosen by the stopping strategy for different values of $\alpha$ in Eq. \eqref{eq:earlyStopCriteria}. Finally, Figures \ref{fig:loss_picnn1} and \ref{fig:loss_picnn2} show the loss evolution for PI-ICNN-based approaches. These approaches do not exhibit severe overfitting, and the validation loss plateaus rather than increasing. We hypothesize that this is due to the regularizing effect of the permutation-invariance.

\setcounter{table}{0}
\section{$\text{PI-ICNN}_2$ variants} \label{app:PIICNN2_variants}
Here we present two alternative designs for $\text{PI-ICNN}_2$ networks. These methods either differ in the architecture or the network inputs as detailed below.

\paragraph{Variant 1}
This variant has a simpler architecture for $\text{PI-ICNN}_2$ compared to the one proposed in Eq. \eqref{Eq:PI-ICNN2}. We write it as

\begin{equation}
    h({\hat{\bm{s}}}) = \mathcal{N}_{MC} \left(\left[\frac{1}{6} \sum_{i=1}^6  \mathcal{N}_{C} (\hat{s}_i)^p \right]^\frac{1}{p}  \right) ,
\end{equation}

where $\hat{\bm{s}} = \{\bm{s}', \bm{s}'' \}$. In contrast to the network proposed in the main text, this variant trains only a single input convex neural network for all six inputs. However, in doing so, we are over-constraining the network because this way we are enforcing permutation invariance with respect to all six inputs at the same time.

\paragraph{Variant 2}
This second variant assumes a similar neural network architecture to variant 1 but differs in its inputs. Instead of directly using the transformed stress components as inputs to the network, we use the absolute differences of these components, similar to what we see in Yld2004-18p. However, since the absolute function is already applied, we now need both networks to be convex and monotonically nondecreasing. The output is obtained via
\begin{equation}
    h(\tilde{\bm{s}}) = \bar{\mathcal{N}}_{MC} \left(\left[\frac{1}{9} \sum_{i=1}^9  \bar{\bar{\mathcal{N}}}_{MC} (\tilde{s}_i)^p \right]^\frac{1}{p}  \right) ,
\end{equation}

where the bars on $\mathcal{N}_{MC}$ are used to distinguish separate networks and we define $\tilde{\bm{s}}$ as

\begin{equation}
\tilde{\bm{s}}
=
\{
\lvert {s}'_{1}-{s}''_{1}\rvert, \
\lvert {s}'_{1}-{s}''_{2}\rvert, \
\lvert {s}'_{1}-{s}''_{3}\rvert, \
\lvert {s}'_{2}-{s}''_{1}\rvert, \
\lvert {s}'_{2}-{s}''_{2}\rvert, \
\lvert {s}'_{2}-{s}''_{3}\rvert, \
\lvert {s}'_{3}-{s}''_{1}\rvert, \
\lvert {s}'_{3}-{s}''_{2}\rvert, \
\lvert {s}'_{3}-{s}''_{3}\rvert
\}.
\end{equation}

We evaluate these two variants on the nine datasets presented in Table \ref{tab:dataset_splits} by choosing $\alpha=10$ in Eq. \eqref{eq:earlyStopCriteria} to define the stopping epoch. Both methods have 243 trainable parameters. The results are reported in Tables \ref{tab:PI-ICNN2-method1} and \ref{tab:PI-ICNN2-method2}. The number of trainable parameters for these methods is comparable with those reported in Table \ref{tab:no_parameters}. We can therefore compare the results of these variants with those reported for $\text{PI-ICNN}_2$ in Tables \ref{tab:training_lowerParam_alpha10}, \ref{tab:validation_lowerParam_alpha10}, and \ref{tab:test_lowerParam_alpha10}. We can see that both variants perform worse in predicting yield and Lankford ratios. We believe that for variant 1, this can be attributed to the over-constrained nature of the network. Variant 2 subpar performance is possibly due to the use of multiple compositions and the bias introduced in the network due to the choice of absolute differences as inputs.

\begin{table}[h!]
\centering
\caption{Average error metrics for variant 1.}
\label{tab:PI-ICNN2-method1}
\begin{tabular}{lcccc}
\toprule
Dataset & Max $|f|$ & Mean $|f|$ & Max $|\Delta r|$ & Mean $|\Delta r|$ \\
\midrule
Training & 19.80 & 8.76 & 0.03 & 0.02 \\
Validation & 47.40 & 33.36  & 0.22 & 0.15 \\
Testing & 47.30 & 33.10 & 0.89 & 0.57 \\
\bottomrule
\end{tabular}
\end{table}

\begin{table}[h!]
\centering
\caption{Average error metrics for variant 2.}
\label{tab:PI-ICNN2-method2}
\begin{tabular}{lcccc}
\toprule
Dataset & Max $|f|$ & Mean $|f|$ & Max $|\Delta r|$ & Mean $|\Delta r|$ \\
\midrule
Training & 55.37 & 27.36 & 0.34 & 0.09 \\
Validation & 61.04 & 42.61  & 0.25 & 0.17 \\
Testing & 52.61 & 37.33 & 0.86 & 0.53 \\
\bottomrule
\end{tabular}
\end{table}

\section{Performance on test dataset} \label{testDataset_Figures}

\begin{figure}
    \centering
    \includegraphics[width=1.0\linewidth]{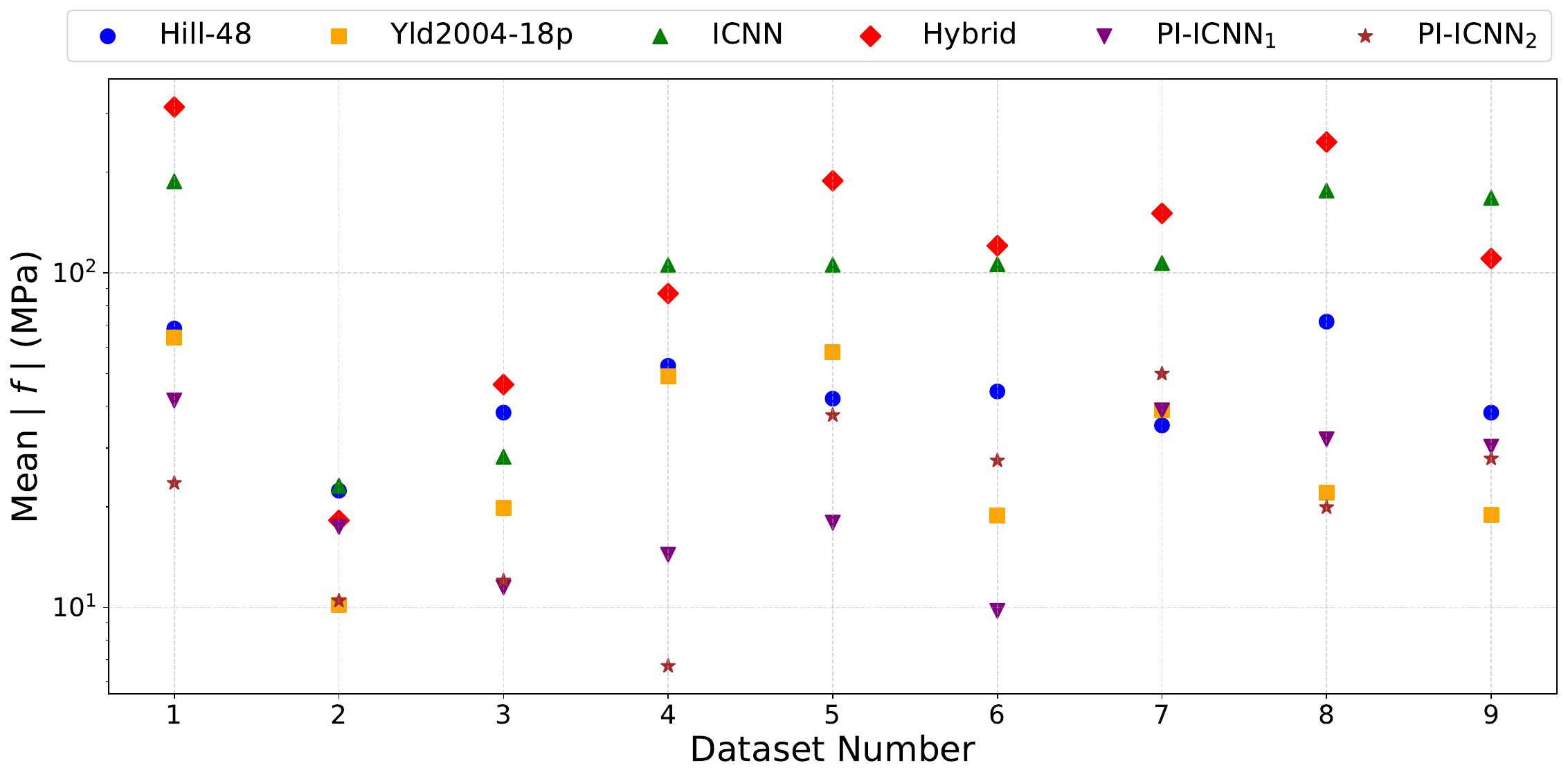}
    \caption{Comparison of each framework's performance in predicting the yield for the test dataset.}
    \label{fig:mean_f_scatter}
\end{figure}

\begin{figure}
    \centering
    \includegraphics[width=1.0\linewidth]{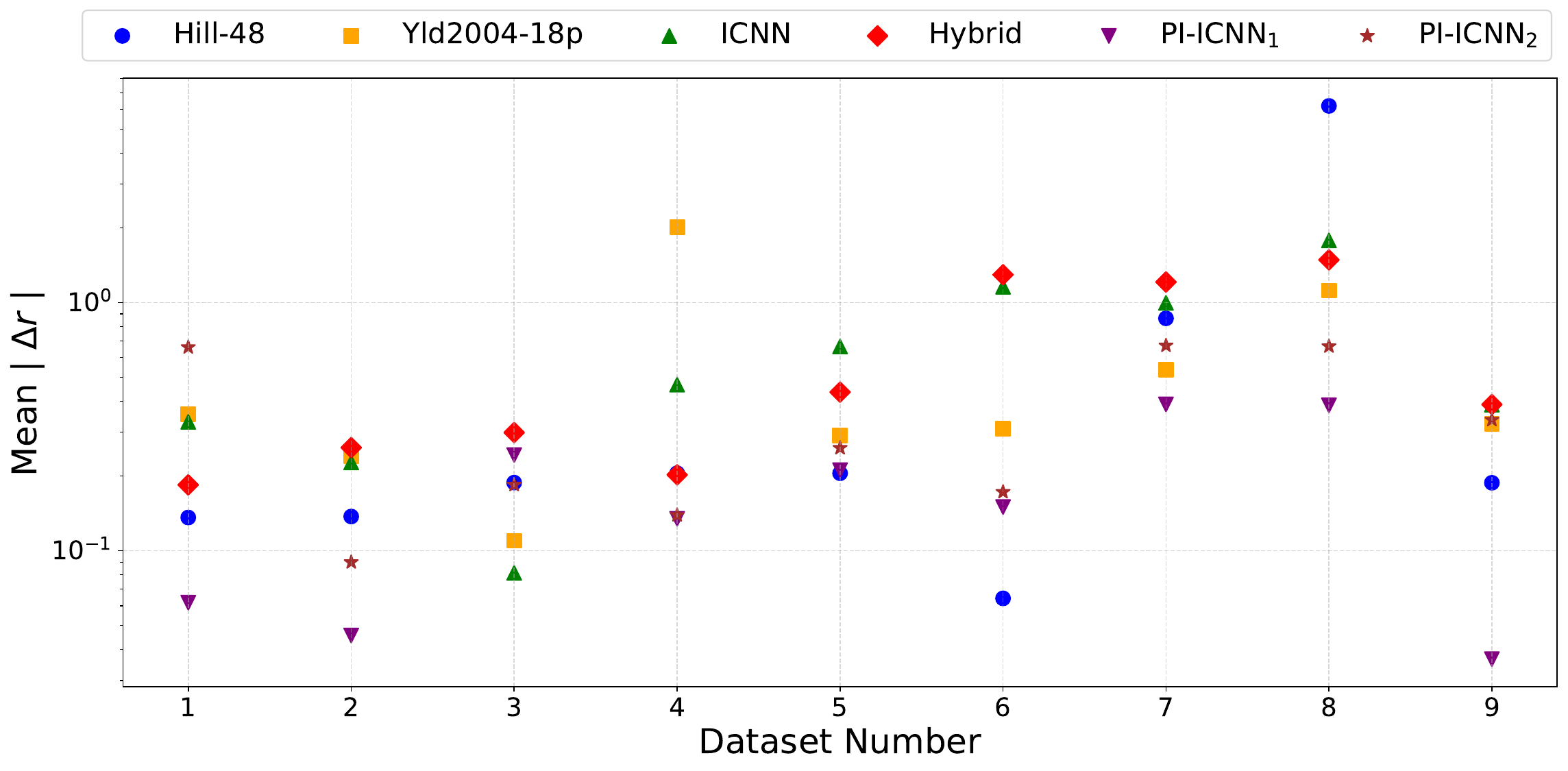}
    \caption{Comparison of each framework's performance in predicting the r-values for the test dataset.}
    \label{fig:mean_dr_scatter}
\end{figure}

In this section, we make a visual comparison between each framework's performance on the (unseen) test data across all nine dataset splits. Figure \ref{fig:mean_f_scatter} shows the mean of the predicted yield for the two test data points across all splits whereas Figure \ref{fig:mean_dr_scatter} shows the mean discrepancy in the predicted r-values. Note the use of a log scale on the y-axis. Although we only present the averaged errors over the nine datasets throughout the text, Figures \ref{fig:mean_f_scatter} and \ref{fig:mean_dr_scatter} show that those averaged values are representative of frameworks' performance on each of the dataset split, barring some exceptions.

\end{document}